\documentclass[aps,prl,twocolumn,superscriptaddress, amsmath,amssymb,nofootinbib]{revtex4-2}

\usepackage{diagbox}
\usepackage{ragged2e}
\usepackage{array}
\usepackage{booktabs}
\usepackage{graphicx}
\usepackage{dcolumn}
\usepackage{bm}
\usepackage[colorlinks, linkcolor=black, anchorcolor=black, citecolor=black]{hyperref}

\providecommand{\enquote}[1]{#1}

\begin{document}

\preprint{APS/123-QED}

\title{Unified Description for Reentrance and $T_c$ Enhancement in Ferromagnetic Superconductors}

\author{Xusheng Wang}
\email{wang-xs23@mails.tsinghua.edu.cn}
\affiliation{%
State Key Laboratory of Low-Dimensional Quantum Physics, Department of Physics, Tsinghua University, Beijing 100084, China}
\author{Lianyi He}
\affiliation{%
State Key Laboratory of Low-Dimensional Quantum Physics, Department of Physics, Tsinghua University, Beijing 100084, China}
\author{Shuaihua Ji}
\affiliation{%
State Key Laboratory of Low-Dimensional Quantum Physics, Department of Physics, Tsinghua University, Beijing 100084, China}
\affiliation{%
Frontier Science Center for Quantum Information, Beijing 100084, China}

\date{\today}

\begin{abstract}
Ferromagnetic superconductors, where ferromagnetism and superconductivity coexist despite their antagonism, exhibit strikingly diverse behaviors. Depending on the interplay between the ferromagnetic exchange field and the superconducting condensate, superconductivity may vary with temperature monotonically, non-monotonically, or even reentrantly, all of which can be tuned by external magnetic fields. Here, we present a unified theoretical framework and construct comprehensive phase diagrams that capture these regimes and predict new phenomena, including double-reentrant superconductivity. We further demonstrate how magnetic fields drive systematic shifts between distinct temperature-dependent behaviors, thereby explaining recent experimental results and predicting new transitions. In addition, we show that the field-induced enhancement of superconductivity, previously observed as a slight increase in $T_c$, can under certain conditions become dramatic.

\end{abstract}

\maketitle


Ferromagnetism, featured with a large exchange field, typically suppresses superconductivity once the field exceeds the paramagnetic Pauli limit $H_p$ \cite{clogston1962upper,chandrasekhar1962note,sarma1963influence,maki1966effect}. Nevertheless, intrinsic ferromagnetic superconductors (FSC) have been discovered in heavy-fermion uranium and europium compounds \cite{saxena2000superconductivity,aoki2001coexistence,hardy2005p,huy2007superconductivity,ran2019nearly,jiang2009superconductivity,iida2019coexisting,tran2024reentrant}, where the coexistence of these antagonistic orders gives rise to exotic reentrant phenomena and has become a central focus of recent research. Such reentrant superconductivity can be broadly divided into field-induced and temperature-induced types \cite{meul1984observation,ran2019extreme,helm2024field}. The former is often explained by the Jaccarino-Peter effect \cite{jaccarino1962ultra}, where an external magnetic field compensates the exchange field, thereby restoring superconductivity. The latter, in contrast, is only qualitatively attributed to magnetic phase transitions \cite{mineev2015reentrant,mineev2017superconductivity,aoki2019review}. Experimentally, temperature-dependent behaviors of FSC fall into three distinct categories: (i) monotonic, where superconductivity appears at low temperature and vanishes at higher temperatures \cite{eisaki1994competition,huy2007superconductivity}; (ii) non-monotonic, where superconductivity survives only in an intermediate temperature range \cite{ishikawa1977destruction,fertig1977destruction,tran2024reentrant}; and (iii) reentrant, where superconductivity persists at low temperatures, disappears at slightly higher temperatures, and reemerges upon further heating \cite{eisaki1994competition,paramanik2013reentrant}. Despite extensive efforts, a unified theory that accounts for these diverse behaviors remains absent.

In this Letter, we analyze the superconducting state of FSC by incorporating a temperature-dependent exchange field and constructing the phase diagram in the exchange-field-temperature plane. This framework successfully reproduces all experimentally observed types of temperature-dependent behaviors and further predicts two previously unreported reentrant regimes. Moreover, we demonstrate that these behaviors are not fixed but can be transformed into one another by tuning an external compensating magnetic field, providing a natural explanation for several key experimental observations. Most strikingly, our analysis reveals that the superconducting transition temperature $T_c$, which is only slightly enhanced by magnetic fields in previous experiments \cite{jeffrey2011enhancement,tran2012tuning,hu2025lithium}, can under special conditions increase by up to a factor of five.

For a FSC, the ferromagnetic contribution can be effectively described by an exchange field $h_{ex}$ under the assumption of weak coupling between magnetism and superconductivity \cite{stoner1938collective,jaccarino1962ultra}. The Hamiltonian of the system is then written as
{\small
\begin{align}
\hat{H} &= \sum_{\mathbf{k},\sigma} \xi(\mathbf{k}) 
\hat{\psi}^\dagger_{\sigma}(\mathbf{k}) \hat{\psi}_{\sigma}(\mathbf{k}) 
+ \sum_{\mathbf{k},\sigma} \sigma h_{ex} 
\hat{\psi}^\dagger_{\sigma}(\mathbf{k}) \hat{\psi}_{\sigma}(\mathbf{k}) \notag \\
&\quad + \sum_{\mathbf{k},\mathbf{k}'} U(\mathbf{k},\mathbf{k}')
\hat{\psi}^\dagger_{\uparrow}(\mathbf{k}) 
\hat{\psi}^\dagger_{\downarrow}(-\mathbf{k})
\hat{\psi}_{\downarrow}(-\mathbf{k}') 
\hat{\psi}_{\uparrow}(\mathbf{k}'),
\end{align}
}

\noindent where the second and third terms describe the ferromagnetic exchange field and the superconducting pairing interaction, respectively. Within the mean-field approximation and assuming an isotropic order parameter, the thermodynamic potential $\Omega$ takes the form (see supplementary material (SM) for details)

{\small
\begin{align}
\Omega &= N \Delta^2 \left( \ln \frac{\Delta}{\Delta_0} - \frac{1}{2} \right) - \frac{2N}{\beta} \int_0^{\infty} d\xi \, \notag \\
&\qquad \ln \Biggl( 1 + 2 \cosh(\beta h_{ex}) e^{-\beta \sqrt{\xi^2 + \Delta^2}}
+ e^{-2\beta \sqrt{\xi^2 + \Delta^2}} \Biggr),
\end{align}
}

\noindent where $N$ is the density of states, $\Delta$ the superconducting order parameter with $\Delta_0$ the value at $h_{ex}=T=0$, and $\beta = 1/k_B T$. Minimizing $\Omega$ yields $\Delta(h_{ex},T)$, from which the phase diagram can be mapped, with the superconducting and normal states separated by the condition $\Delta(h_{ex},T)=0$. Importantly, the exchange field in a ferromagnet itself exhibits a strong temperature dependence, which can be generally described as \cite{kuz2006factors,kuz2006temperature}

\begin{table*}[]
\centering 
\renewcommand{\arraystretch}{1.4}
\caption{Classification of temperature-dependent superconductivity in FSC, based on the topology between the exchange-field curve $h_{ex}(T)$ and the superconducting phase-transition line.}
\begin{tabular}{l@{\hskip 1.7cm}c@{\hskip 1.7cm}c@{\hskip 1.7cm}c}
\hline
Type of behavior & Monotonic & Non-monotonic & Reentrant \\ 
$h_{ex}(0)$ vs.~$H_{p}$ & $h_{ex}(0) < H_{p}$ & $h_{ex}(0) > H_{p}$ & $h_{ex}(0) < H_{p}$ \\ 
Number of intersections $M$ & $M = 0$ or $1$ & $M = 1$ or $2$ & $M = 2$ or $3$  \\ 
Possible topological classes & 3  & 3 & 3  \\ 
Representative material & $\mathrm{LuNi_2B_2C}$ \cite{eisaki1994competition} & $\mathrm{ErRh_4B_4}$ \cite{crespo2009evolution} & $\mathrm{HoNi_2B_2C}$ \cite{eisaki1994competition}\\ \hline
\end{tabular}\label{class}
\end{table*}

\begin{figure}[ht]
  \centering
  \includegraphics[width=0.95\linewidth]{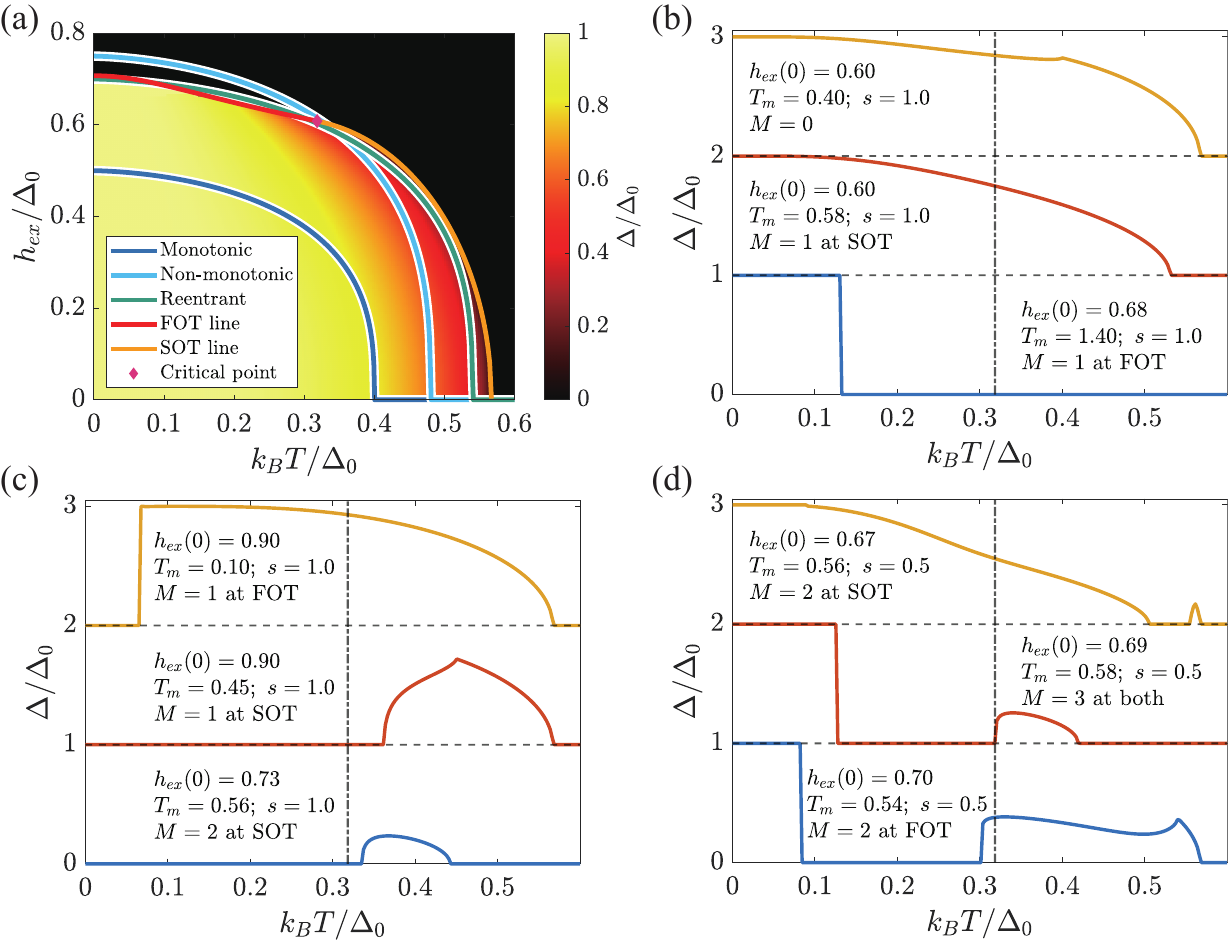}
  \caption{Phase diagrams and three experimentally observed temperature-related behaviors. (a) $h_{ex}$-$T$ phase diagram showing superconducting and normal states separated by first-order (FOT, red) and second-order (SOT, orange) phase-transition lines. The color scale indicates the superconducting gap amplitude. Example $h_{ex}(T)$ curves (light blue, cyan, green) illustrate monotonic, non-monotonic, and reentrant behaviors. (b-d) Calculated $\Delta(T)$ curves for the monotonic (b), non-monotonic (c), and reentrant (d) types, each further classified by the topological relation between the $h_{ex}(T)$ curve and the phase-transition lines (curves offset vertically by 1 for clarity). The dashed-dotted black line marks the critical points (similarly in Fig.~\ref{phase_diagram_parameter_space}(d)). $h_{ex}$ and $T_m$ are both expressed in units of $\Delta_0$, as in the following discussion.}
  \label{three_type_observed}
\end{figure}

{\small
\begin{equation}\label{T_exchange}
\frac{h_{ex}(T)}{h_{ex}(0)} =
\left[ 1 - s \left( \frac{T}{T_m} \right)^{3/2}
- (1-s) \left( \frac{T}{T_m} \right)^{5/2} \right]^{1/3},
\end{equation}
}

\noindent where $0 < s < 5/2$ is an empirical parameter determined by material properties, and $T_m$ is the Curie temperature. If $h_{ex}$ is treated as a temperature-independent parameter, the $h_{ex}$-$T$ phase diagram is well established \cite{sarma1963influence}. By plotting the $h_{ex}(T)$ curve on this phase diagram, finite temperature behavior of superconductivity can be distinguished through its intersection with the phase transition boundary. Remarkably, the three experimentally observed types of temperature dependence—monotonic, non-monotonic, and reentrant—can be uniformly understood within this framework by tuning the ferromagnetic parameters $s$, $h_{ex}(0)$, and $T_m$, as shown in Fig.~\ref{three_type_observed}(a).

Each superconducting type can be classified by the topological relation between the temperature-dependent exchange-field curve and the superconducting phase-transition line, as summarized in Table~\ref{class}. For monotonic type, the condition $h_{ex}(0)<H_p$ holds and the two curves intersect at no more than one point [Fig.~\ref{three_type_observed}(b)]. The non-monotonic type arises when $h_{ex}(0)>H_p$ and one or two intersections occur [Fig.~\ref{three_type_observed}(c)]. Most intriguingly, temperature-induced reentrant superconductivity emerges when $h_{ex}(0)<H_p$ and the curves intersect at two or three points [Fig.~\ref{three_type_observed}(d)]. The calculated temperature dependence of the gap amplitude reproduces experimental results from resistance measurements \cite{eisaki1994competition,huy2007superconductivity,ishikawa1977destruction,fertig1977destruction,tran2024reentrant,jiang2009superconductivity,paramanik2013reentrant} and tunneling spectroscopy \cite{crespo2009evolution} (see SM for detailed comparisons). Notably, our model predicts unconventional non-monotonic gap variations with temperature [Figs.\ref{three_type_observed}(b-d)], which have not yet been observed due to the lack of systematic tunneling studies, but may be detectable in $\mathrm{HoNi_2B_2C}$ or $\mathrm{ErNi_2B_2C}$ \cite{eisaki1994competition}. Furthermore, the order of the superconducting transition itself depends on temperature: it is first order at low temperatures, where the specific heat diverges, and second order at higher temperatures, where the specific heat shows a discontinuity. These two thermodynamic signatures are consistent with experimental observations on $\mathrm{EuRh_4B_4}$ and $\mathrm{LuRh_4B_4}$ \cite{woolf1979superconducting,fertig1977destruction} (see SM).

\begin{figure}[ht]
  \centering
  \includegraphics[width=1.0\linewidth]{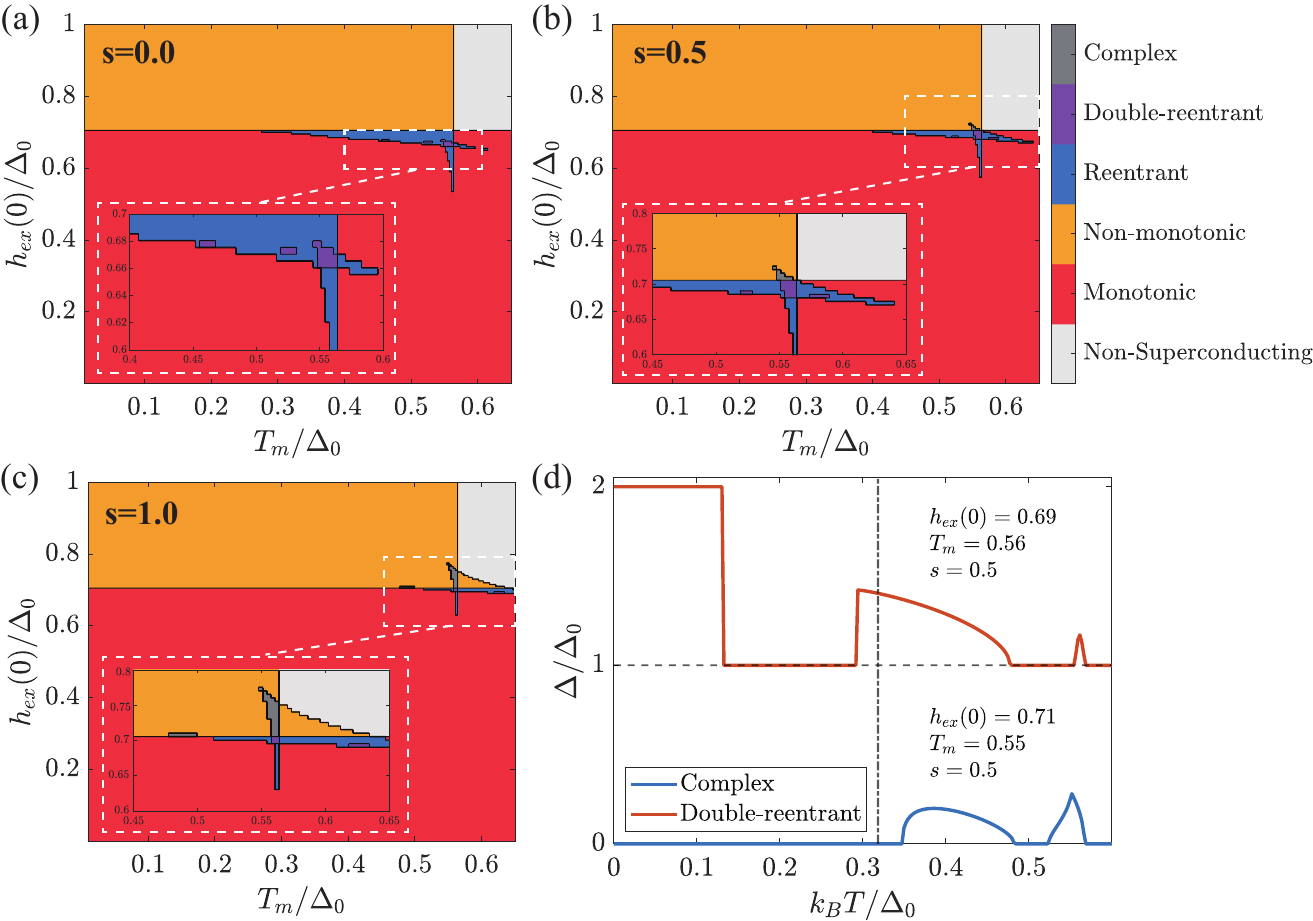}
  \caption{Phase diagrams in the $h_{ex}(0)$-$T_m$ plane revealing novel behaviors. (a-c) Calculated phase diagrams for $s=0$ (a), $0.5$ (b), and $1.0$ (c), with different superconducting regimes classified by color (see color scale). Insets highlight the narrow reentrant regions. (d) Representative $\Delta(T)$ curves for the predicted double-reentrant and complex behaviors.}
  \label{phase_diagram_parameter_space}
\end{figure}

To further elucidate the parameter dependence of temperature-related behaviors, we numerically calculated the phase diagrams of $h_{ex}(0)$ versus $T_m$ for fixed values of $s$ [Figs.~\ref{phase_diagram_parameter_space}(a)-\ref{phase_diagram_parameter_space}(c)]. In the calculation, the superconducting behaviors were classified as follows. Superconductivity was considered terminated when $\Delta < 10^{-4}\Delta_0$ and it was denoted as `N', otherwise labeled as `S'. Compressing consecutive identical strings yields compact sequences (e.g., `$\mathrm{S},\mathrm{N},\cdots$' or `$\mathrm{N},\mathrm{S},\cdots$'), which uniquely distinguish the different types of behavior. In this representation, monotonic, non-monotonic, and reentrant superconductivity correspond to `S-N', `N-S-N' and `S-N-S-N' respectively. Beyond these, additional categories emerge, such as `N' for a fully nonsuperconducting state, and more complex sequences like `N-S-N-S-N' and `S-N-S-N-S-N' which we identify as novel complex and double-reentrant behaviors that have not yet been experimentally observed.

\begin{table*}[]
\centering
\renewcommand{\arraystretch}{1.4}
\setlength{\tabcolsep}{8pt}
\caption{Magnetic-field-induced transformations of temperature-dependent superconducting behaviors for $s=0.5$ with representative parameters. The first row and column denote the initial and final phases without and with magnetic field, respectively. In each entry, the left value indicates the theoretical possibility (evaluated by phase-diagram coverage), and the right value lists available experimental results.}
\scalebox{0.66}{
\begin{tabular}{l@{\hskip 0.4cm}c@{\hskip 0.4cm}c@{\hskip 0.4cm}c@{\hskip 0.4cm}c@{\hskip 0.4cm}c@{\hskip 0.4cm}c}
\hline
\diagbox{Final}{Initial} & Non-Superconducting & Monotonic & Non-monotonic & Reentrant & Double-reentrant & Complex \\ \hline
Monotonic      & Probably/$\mathrm{UTe_2}$ \cite{frank2024orphan} & -- & Probably/$\mathrm{Eu_{x}Ca_{1-x}(Fe_{y}Co_{1-y})_2As_2}$ \cite{tran2024reentrant} & Probably/($\mathrm{HoNi_2B_2C}$ \cite{eisaki1994competition}) & Probably/Not yet & Probably/Not yet \\
Non-monotonic  & Unlikely/Not yet & Probably/($\mathrm{TmNi_2B_2C}$ \cite{eisaki1994competition}) & -- & Probably/Not yet & Probably/Not yet & Probably/Not yet \\
Reentrant      & Possibly/Not yet & Possibly/($\mathrm{HoNi_2B_2C}$ \cite{anisotropic1996}) & Possibly/Not yet & -- & Probably/Not yet & Probably/Not yet \\
Double-reentrant & Impossibly/No & Unlikely/Not yet & Unlikely/Not yet & Possibly/Not yet & -- & Probably/Not yet \\
Complex        & Impossibly/No & Impossibly/No & Unlikely/Not yet & Possibly/Not yet & Unlikely/Not yet & -- \\
\hline
\end{tabular}}
\label{tab:phase_switch}
\end{table*}

As shown in the phase diagrams, at least five distinct temperature-related behaviors appear for all values of $s$, demonstrating the universality of the results derived from Eq.~(\ref{T_exchange}). Among them, the monotonic and non-monotonic regimes occupy the largest areas of parameter space, while the reentrant regime is comparatively small. This naturally explains why most FSC exhibit monotonic or non-monotonic behavior, whereas reentrant behavior is observed only in a few cases. Increasing $s$ leads to a progressive shrinkage of the reentrant region, while the double-reentrant and complex phases gradually expand. Nevertheless, these latter regimes remain confined to very narrow regions of parameter space, consistent with the fact that they have not yet been experimentally identified.

It is also instructive to examine the intriguing double-reentrant and complex regimes. As illustrated by the phase diagram and $\Delta(T)$ curves in Fig.~\ref{phase_diagram_parameter_space}(d), these behaviors can again be understood in terms of the topological relation between the $h_{ex}(T)$ curve and the phase-transition line. Double-reentrant superconductivity occurs when $h_{ex}(0)<H_p$ and four or five intersections appear, while the complex regime arises for $h_{ex}(0)>H_p$ with three or four intersections, resembling the reentrant case. Since temperature-dependent magnetic transitions can further modify the exchange field \cite{aoki2014superconductivity,aoki2019review,ram2023multiple,zhou2024multiple}, even richer temperature-dependent superconducting behaviors may emerge from additional intersection points or shifts in the relative magnitude of $h_{ex}(0)/H_p$, a possibility that lies beyond the scope of this Letter.

According to the discussion above, the divergent temperature behaviors originate from the distinct forms of the exchange field relation. Notably, an effective exchange field can also be introduced by applying an external magnetic field $h$, thereby providing a means to tune ferromagnetic superconductivity. Supported by experimental observations \cite{meul1984observation,ran2019extreme,helm2024field}, the orbital contribution in many heavy-fermion ferromagnets is insignificant, so the external field primarily acts as an additional exchange field \cite{myNote}. Since the intrinsic ferromagnetic exchange field has a fixed direction but a temperature-dependent magnitude, adjusting the orientation and strength of the external field offers a direct route to modify the overall temperature dependence. The total exchange field is then given by

{\small
\begin{equation}\label{exchang_with_external}
h_t(T) = \sqrt{(h_{ex}(T) + h\cos\theta)^2 + h^2\sin^2\theta},
\end{equation}
}

\noindent where $\theta$ is the angle between the applied field $h$ and the intrinsic exchange field $h_{ex}(T)$. For fixed values of $h_{ex}(0)$, $T_m$, and $s$, the resulting $h$-$\theta$ phase diagram can be systematically classified using the same string-compression method introduced above.

To illustrate the tunability of temperature-dependent superconductivity under magnetic fields, we focused on the representative case of $s=0.5$. In this regime, all phases can in principle emerge for suitable choices of $h_{ex}(0)$ and $T_m$. To track their evolution, we selected representative parameter sets corresponding to each phase at $h=0$ and mapped the associated $h$-$\theta$ phase diagrams. Given the large number of possible phases, we restricted the main discussion to the monotonic case, which is of particular experimental relevance, while the full analysis of other phases is provided in the SM. A summary of both theoretical and experimental results is presented in Table~\ref{tab:phase_switch}. As shown, our framework not only reproduces previously observed field-induced transformations, but also predicts several additional, more exotic phase-switching processes.

\begin{figure}[ht]
  \centering
  \includegraphics[width=1.0\linewidth]{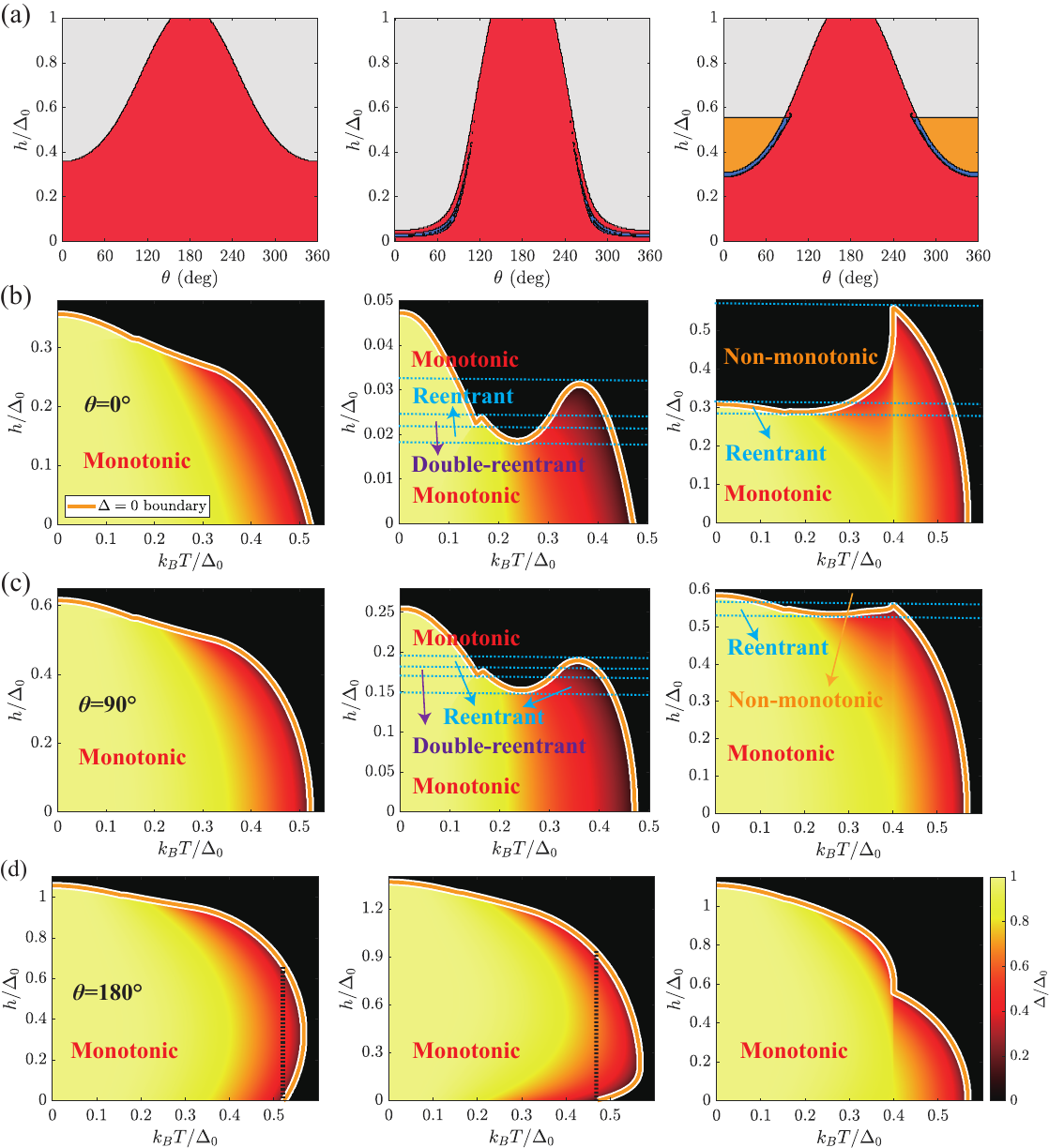}
  \caption{Magnetic field induced transformations of monotonic superconductivity.  
  (a) $h$-$\theta$ phase diagrams for three types of monotonic behavior, classified by color (color scale as in Fig.~\ref{phase_diagram_parameter_space}).  
  (b-d) $\Delta(T,h)$ maps for each case along $\theta=0^\circ$ (b), $90^\circ$ (c), and $180^\circ$ (d). Distinct temperature-dependent behaviors at different $h$ are separated by dashed light-blue lines, with gap amplitude denoted by the color scale (lower right). The dashed line in (d) highlights field-induced enhancement of superconductivity. Parameter sets: first column, $M=1$ case ($h_{ex}(0)=0.35$, $T_m=1.40$, $s=0.50$); second column, $M=1$ case near the $M=1/M=0$ boundary ($h_{ex}(0)=0.66$, $T_m=0.58$, $s=0.50$); third column, $M=0$ case ($h_{ex}(0)=0.40$, $T_m=0.40$, $s=0.50$).}
  \label{phase_diagram_magnetic_field}
\end{figure}

The monotonic behaviors can be further classified into three types [Fig.~\ref{three_type_observed}(a)]. Under an applied magnetic field, however, we find that the transformation behavior is determined almost solely by the number of intersection points $M$ (see SM for details). Specifically, $M=0$ and $M=1$ correspond to the conditions $T_m<T_{c0}$ ($T_{c0}\approx 0.57\Delta_0$) and $T_m>T_{c0}$, respectively, indicating that $T_m=T_{c0}$ serves as the boundary between the two topological classes. Our analysis further reveals that states located near this boundary exhibit mixed features of both regimes, giving rise to a distinct form of transformation. To capture these effects, we analyzed the magnetic response for monotonic behaviors with $M=0$, near the $T_m=T_{c0}$ boundary, and with $M=1$, as summarized in the phase diagram of Fig.~\ref{phase_diagram_magnetic_field}(a).

The $h$-$\theta$ phase diagrams highlight that the direction of the applied magnetic field often overlooked in experiments \cite{eisaki1994competition,ishikawa1977destruction}, is crucial for tuning temperature-dependent behaviors. To illustrate this, we present $\Delta(T,h)$ maps at fixed field orientations [Figs.~\ref{phase_diagram_magnetic_field}(b)-(d)]. For the $M=1$ case, the system remains monotonic for all field strengths and directions, while the critical field increases significantly as $\theta$ varies from $0^\circ$ to $180^\circ$. Near the $T_m=T_{c0}$ boundary, however, more complex responses emerge. At $\theta=0^\circ$ or $90^\circ$, the behavior evolves sequentially from monotonic to reentrant, to double-reentrant, back to reentrant, and finally to monotonic as the field strength increases; the critical field is notably larger at $\theta=90^\circ$. Corresponding transitions between monotonic-reentrant-monotonic superconductivity at $\theta=0^\circ$ or $90^\circ$ have been observed in $\mathrm{HoNi_2B_2C}$ \cite{anisotropic1996}, though the predicted double-reentrant phase has not yet been detected, likely due to its narrow parameter range and experimental field resolution. For the $M=0$ case, monotonic superconductivity can evolve into a narrow reentrant region and then into non-monotonic behavior along $\theta=0^\circ$ or $90^\circ$, consistent with monotonic-to-non-monotonic transitions reported in $\mathrm{TmNi_2B_2C}$ \cite{eisaki1994competition}. In contrast, no transformation occurs along $\theta=180^\circ$, where the system remains monotonic. Notably, in the $M=1$ regime at $\theta=180^\circ$, we also find a slight enhancement of $T_c$ induced by the magnetic field [Fig.~\ref{phase_diagram_magnetic_field}(d)].

\begin{figure}[ht]
  \centering
  \includegraphics[width=1.0\linewidth]{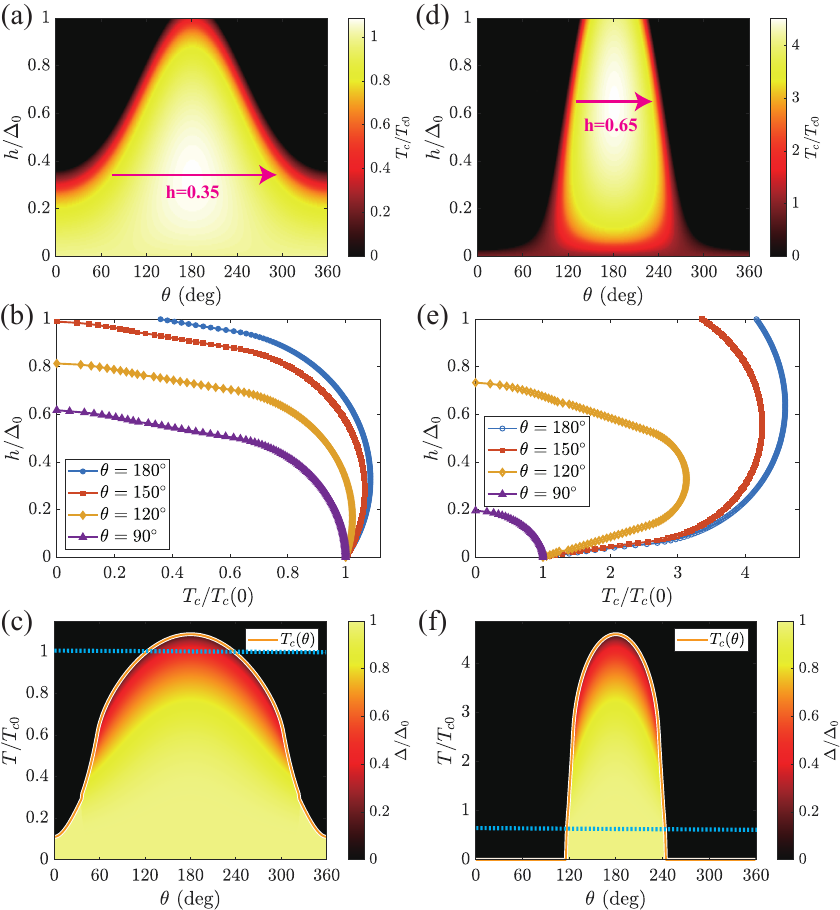}
  \caption{Magnetic-field-induced enhancement of superconductivity.  
  (a,d) $T_c(h,\theta)$ maps for monotonic behaviors with intersections at the second-order (a-c) and first-order (d-f) transition regimes, respectively.  
  (b,e) Field dependence of $T_c$ along $\theta=90^\circ$, $120^\circ$, $150^\circ$, and $180^\circ$ for the two cases.  
  (c,f) $\Delta(T,\theta)$ maps at fixed $h$ along the pink lines in (a,d). Blonde curves trace the field-enhanced $T_c$ as a function of $\theta$, while light-blue lines denote the original zero-field $T_c(0)$. Parameter sets: first column, $h_{ex}(0)=0.35$, $T_m=1.40$ and $s=0.50$; second column, $h_{ex}(0)=0.68$, $T_m=1.40$ and $s=0.50$.}
  \label{enhanced_superconductivity}
\end{figure}

Since monotonic behavior with $M=1$ can occur in two distinct forms, we analyzed the field-induced $T_{c0}$ enhancements for both cases in detail. To capture the response to field strength and orientation, the corresponding $T_c(h,\theta)$ maps are shown in Figs.~\ref{enhanced_superconductivity}(a) and \ref{enhanced_superconductivity}(d).

For intersections at the second-order transition regime, only a slight enhancement of $T_{c}$ is observed in the mapping, whereas for intersections at the first-order regime, requiring $T_m>T_{c0}$ and $h_{ex}\lesssim H_p$, a dramatic $T_c$ increase up to nearly fivefold emerges. To further clarify the field dependence, $T_c(h)$ curves along $\theta=90^\circ \rightarrow 180^\circ$ are presented in Figs.\ref{enhanced_superconductivity}(b) and \ref{enhanced_superconductivity}(e). Both reveal that $T_c$ enhancement occurs for $90^\circ<\theta<270^\circ$, consistent with the interpretation that the external magnetic field compensates the intrinsic exchange field. While modest field-induced $T_c$ enhancements have been reported in lithium-intercalated FeSe and related systems \cite{jeffrey2011enhancement,tran2012tuning,hu2025lithium}, the tremendous enhancement predicted here remains to be experimentally realized. Finally, the angular sensitivity of the enhancement was analyzed by fixing $h$ to maximize $T_c$ in Figs.\ref{enhanced_superconductivity}(a) and \ref{enhanced_superconductivity}(d) and mapping $\Delta(T,\theta)$ [Figs.~\ref{enhanced_superconductivity}(c) and \ref{enhanced_superconductivity}(f)]. Compared with the zero-field case, the enhancement is highly angle-dependent and, for certain parameters (see SM), can be fragile at fixed $h$. In particular, monotonic behavior with strong $T_c$ enhancement exhibits a critical angle beyond which $T_c$ drops sharply from its elevated value.

In summary, we have systematically analyzed the finite temperature superconducting behavior of FSC and their response to magnetic fields, based on a model incorporating a temperature-dependent ferromagnetic exchange field and an isotropic superconducting order parameter. While the model can be generalized by including anisotropic pairing and explicit coupling between ferromagnetism and superconductivity, even in its minimal form it successfully reproduces all experimentally observed temperature-related behaviors \cite{eisaki1994competition,huy2007superconductivity,ishikawa1977destruction,fertig1977destruction,tran2024reentrant,jiang2009superconductivity,paramanik2013reentrant} and classifies them according to the topological relation between the $h_{ex}(T)$ curve and the superconducting phase boundary. Distinct $\Delta(T)$ curves are predicted for each regime [Figs.~\ref{three_type_observed}(b)-(d)], motivating future tunneling spectroscopy in systems such as $\mathrm{HoNi_2B_2C}$ and $\mathrm{ErNi_2B_2C}$, where only one direct gap measurement has been reported to date \cite{crespo2009evolution}. 

By comparing the $h_{ex}(0)$-$T_m$ phase diagrams, we explain why reentrant behavior is rarely observed, while also identifying novel double-reentrant and complex regimes with limited parameter-space coverage. More importantly, we show that magnetic fields provide a powerful knob to transform between different temperature-dependent behaviors, consistent with existing experimental observations and predicting new transitions [Table~\ref{tab:phase_switch}]. Finally, our framework accounts for the recently reported slight magnetic-field-induced enhancement of $T_c$, and further predicts that under specific conditions $T_c$ can be enhanced by up to a factor of five, a phenomenon potentially observable in $\mathrm{UGe_2}$ under high pressure or in $\mathrm{UCoGe}$ \cite{huy2007superconductivity,saxena2000superconductivity}.

These results unify diverse temperature-related phenomena and their magnetic-field responses in FSC, and provide clear guidance for future experimental exploration.

\textit{Acknowledgments.--} We acknowledge Q.-J. Cheng, Y. Liu, X.-Y. Zhou, Q. Zhu, X.-Y. Shi, C.-Y. Hu and M. Shu for stimulating discussions. This work is financially supported by Innovation Program for Quantum Science and Technology (Grants No. 2023ZD0300500) and the National Natural Science Foundation of China (Grants No. 11427903, 52388201 and 12375136).\\


\nocite{*}

\providecommand{\noopsort}[1]{}\providecommand{\singleletter}[1]{#1}%
%

\pagebreak
\appendix

\onecolumngrid   
\setcounter{figure}{0}
\renewcommand*{\thefigure}{S\arabic{figure}}

\setcounter{equation}{0}
\renewcommand{\theequation}{S\arabic{equation}}

\begin{widetext}
\section{\Large{Supplementary Material}}
\section{I. Thermodynamic Potential derivation and calculation details}

Considering an isotropic pairing interaction, the Hamiltonian in the main text can be rewritten as  

\begin{align}
\hat{H} &= \sum_{\mathbf{k},\sigma} \xi_\sigma(\mathbf{k}) 
c^\dagger_{\sigma}(\mathbf{k}) c_{\sigma}(\mathbf{k}) 
- \sum_{\mathbf{k},\mathbf{k}'} g
c^\dagger_{\uparrow}(\mathbf{k}) 
c^\dagger_{\downarrow}(-\mathbf{k})
c_{\downarrow}(-\mathbf{k}') 
c_{\uparrow}(\mathbf{k}'), \quad \xi_\sigma(\mathbf{k}) = \xi({\mathbf{k}}) + \sigma h_{ex},
\label{hamiltonian}
\end{align}

\noindent where $g$ is the coupling strength. The superconducting order parameter is defined as $\Delta=g\langle c_{\downarrow}(-\mathbf{k}') 
c_{\uparrow}(\mathbf{k}') \rangle$, which can be chosen real. Applying the mean-field approximation, the Hamiltonian (\ref{hamiltonian}) simplifies to

\begin{align}
\hat{H} &= \frac{\Delta^2}{g} + \sum_{\mathbf{k}}\begin{pmatrix}
c^\dagger_{\uparrow}(\mathbf{k}) & c^\dagger_{\downarrow}(-\mathbf{k})
\end{pmatrix}
\begin{pmatrix}
\xi_{\uparrow}(\mathbf{k}) & \Delta \\
\Delta & -\xi_{\downarrow}(-\mathbf{k})
\end{pmatrix}
\begin{pmatrix}
c^\dagger_{\uparrow}(\mathbf{k}) \\
c^\dagger_{\downarrow}(-\mathbf{k})
\end{pmatrix} + 
\sum_{\mathbf{k}} \xi_\downarrow(-\mathbf{k}).
\label{hamiltonian2}
\end{align}

\noindent This Hamiltonian can be diagonalized via the standard Bogoliubov transformation, yielding  

\begin{equation}\label{bogoliubov}
\hat{H} = \frac{\Delta^2}{g} 
+ \sum_{\mathbf{k}} \left( \xi({\mathbf{k}}) - E({\mathbf{k}}) \right) 
+ \sum_{s=\pm,\mathbf{k}} E_s(\mathbf{k}) \,
\alpha_s^\dagger(\mathbf{k}) \alpha_s(\mathbf{k}),
\end{equation}

\noindent where $E_s(\mathbf{k}) = E({\mathbf{k}}) + s h_{ex}$ with $E({\mathbf{k}}) = \sqrt{\Delta^2 + \xi^2(\mathbf{k})}$. Here $\alpha_s$ denotes the Bogoliubov quasiparticles, which obey fermionic statistics. The thermodynamic potential $\Omega$ follows from the partition function, $\Omega = -\tfrac{1}{\beta}\ln \mathcal{Z}$, and can be expressed as \cite{sarma1963influence,he2009stable}

\begin{equation}
\Omega = \frac{\Delta^2}{g} 
+ \sum_{\mathbf{k}} \left( \xi({\mathbf{k}}) - E({\mathbf{k}}) \right) 
- \frac{1}{\beta} \sum_{s=\pm} \sum_{\mathbf{k}} 
\ln \left( 1 + e^{-\beta E_s(\mathbf{k})} \right),
\end{equation}

\noindent Assuming the system is in the weak-coupling regime with a constant density of states $N$ within the interaction window, the thermodynamic potential can be expressed in integral form as

\begin{equation}\label{free_energy}
\Omega = \frac{\Delta^2}{g} 
+ 2N\int_0^{E_c} d\xi 
\left[ \, \xi - \sqrt{\xi^2 +\Delta^2} 
- \frac{1}{\beta} \sum_{s=\pm} 
\ln \left( 1 + e^{-\beta \left( \sqrt{\xi^2 + \Delta^2} + s h_{ex} \right)} \right) \right],
\end{equation}

\noindent where $E_c$ is the cutoff energy. The gap equation follows from the stationary condition $\partial \Omega/\partial \Delta = 0$

\begin{equation}\label{gap_equation}
\Delta \left[ 
\frac{1}{gN} 
- \int_0^{E_c} \mathrm{d}\xi \, 
\frac{1 - \sum_{s} f\!\left( \sqrt{\xi^2 + \Delta^2} + s h_{ex} \right)}
{\sqrt{\xi^2 + \Delta^2}} 
\right] = 0,
\end{equation}

\noindent where $f(E)$ is the Fermi-Dirac distribution function. At zero temperature and in the absence of a magnetic field ($T = h_{ex} = 0$), the integral in Eq.~(\ref{gap_equation}) can be evaluated analytically as

\begin{equation}
\frac{1}{gN} = \int_0^{E_c} \mathrm{d}\xi \, 
\frac{1}{\sqrt{\xi^2 + \Delta_{0}^2}} =\ln\left(\frac{E_c}{\Delta_0}+\sqrt{\left(\frac{E_c}{\Delta_0}\right)^2+1}\right)\approx \ln\left(\frac{2E_c}{\Delta_0}\right). \label{eq:gap0}
\end{equation}

\noindent Using Eq.~(\ref{eq:gap0}) to eliminate the cutoff energy and integrating the temperature-independent term in Eq.~\ref{free_energy}, the thermodynamic potential can be rewritten in the form given in the main text:  

\begin{align}
\Omega = N \Delta^2\left(\ln\frac{\Delta}{\Delta_0} - \frac{1}{2}\right)
- \frac{2N}{\beta}\int_0^\infty \mathrm{d}\xi \ln\left(1 + 2\cosh(\beta h_{ex})e^{-\beta \sqrt{\Delta^2 + \xi^2}} + e^{-2\beta \sqrt{\Delta^2 + \xi^2}}\right). \label{eq:Omega3}
\end{align}

\noindent In Eq.~(\ref{eq:Omega3}), the upper integration limit has been extended from $E_c$ to infinity, which is justified by the convergence of the temperature-dependent term and the assumption that $E_c$ is sufficiently large. To simplify the calculation, we introduce a reduced thermodynamic potential $f(x)$ and define the following dimensionless quantities:  

\begin{equation}
f(x) \equiv \frac{\Omega}{N\Delta_0^2}, \quad x \equiv \frac{\Delta}{\Delta_0}, \quad \delta \equiv \frac{h_{ex}}{\Delta_0}, \quad
t \equiv \frac{k_BT}{\Delta_0}, \quad z \equiv \frac{\xi}{\Delta_0}. \label{eq:dimless1}
\end{equation}

\noindent With these definitions, the reduced thermodynamic potential takes the form

\begin{align}
f(x) = x^2\left(\ln x - \frac{1}{2}\right) - 2t\int_0^\infty \mathrm{d}z \ln\left(1 + 2\cosh(\delta/t)e^{-\sqrt{z^2+x^2}/t} + e^{-2\sqrt{z^2+x^2}/t}\right). \label{eq:fx}
\end{align}

As shown in Eq.~(\ref{eq:fx}), the reduced thermodynamic potential depends only on the reduced order parameter $x \in [0,1]$, for a given exchange field and temperature. In equilibrium, the thermodynamic potential must attain its global minimum. Thus, the self-consistent value of the order parameter, $x^{*} = \Delta/\Delta_0$, is obtained by minimizing $f(x)$. Typical behaviors of $f(x)$ at $h_{ex}=0$ are shown in Fig.~\ref{fx}.

\begin{figure}[ht]
  \centering
  \includegraphics[width=0.9\linewidth]{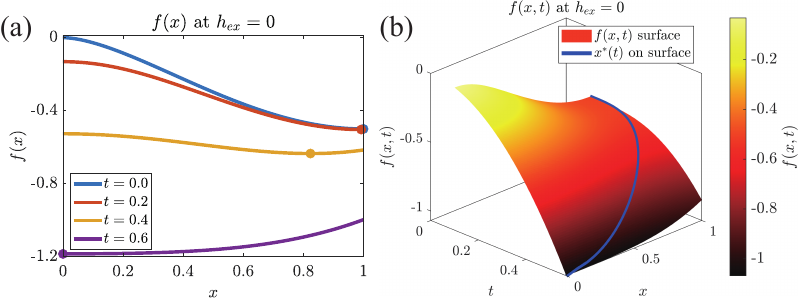}
  \caption{Typical reduced free-energy $f(x)$ at $h_{ex}=0$.  
  (a) $f(x)$ curves for $t=0.0$, $0.2$, $0.4$, and $0.6$, with filled dots marking the global minima.  
  (b) $f(x,t)$ map at $h_{ex}=0$, where the blue line denotes the equilibrium order parameter $x^{*}(t)$.}
  \label{fx}
\end{figure}

As shown in Fig.\ref{fx}(a), the minimum point $x^{*}$ of each $f(x)$ curve corresponds to the equilibrium value of $\Delta$ at that state. It is evident that $\Delta$ decreases gradually with increasing temperature, as expected. By minimizing $f(x)$ at each temperature, the standard BCS $\Delta(T)$ relation is recovered, shown as the blue line in Fig.\ref{fx}(b). The calculation of $\Delta(h_{ex},t)$ can be carried out in the same manner, with the results presented in Fig.~1(a) of the main text.

For convenience, the units of $h_{ex}$ and $T_m$ are expressed in terms of $\Delta_0$ throughout the following discussion.

\newpage
\section{II. Topological-relation-dependent magnetic response}

As discussed in the main text, the magnetic response of a given temperature-dependent superconducting behavior can differ significantly depending on the underlying topological relation—specifically, the number of intersection points $M$ between the phase transition line and the $h_{ex}(T)$ curve. Focusing on the monotonic case, we analyze in greater detail how the magnetic response evolves under different parameter choices.

\begin{figure}[ht]
  \centering
  \includegraphics[width=0.90\linewidth]{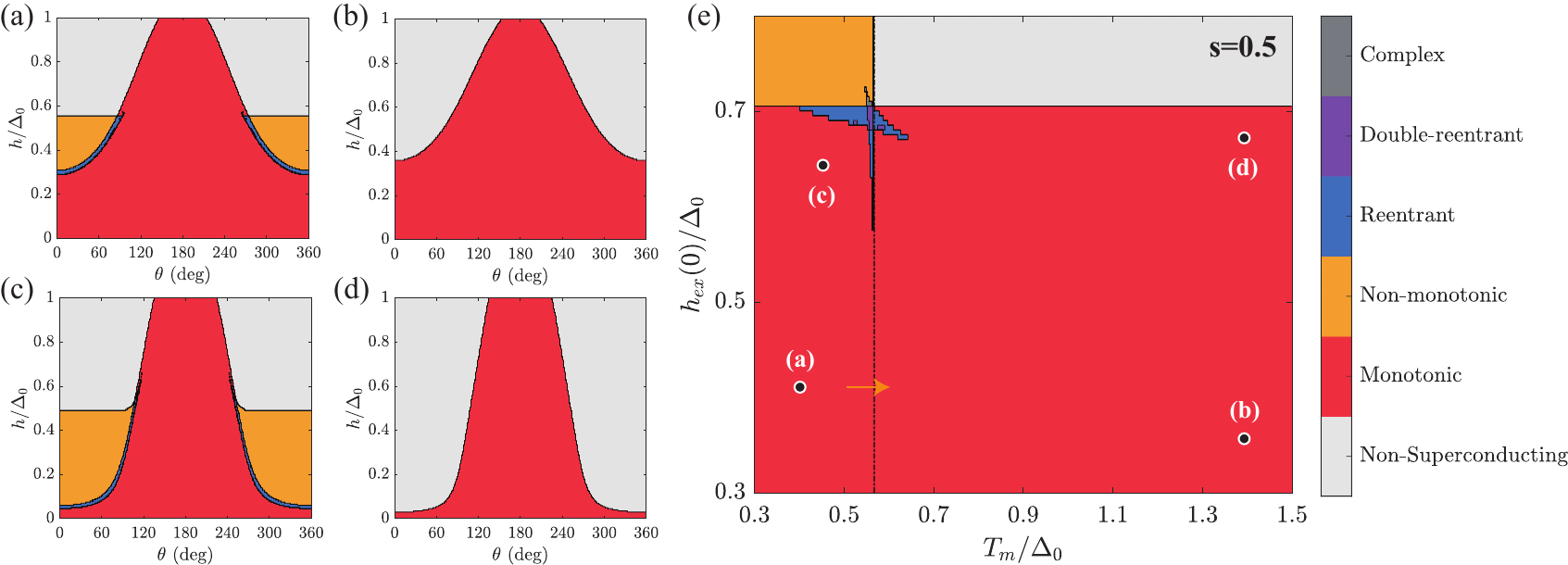}
  \caption{$h$-$\theta$ phase diagrams for the same intersection number $M$ but with different parameters.  
  (a,b) Phase diagrams for (a) $M=0$ ($h_{ex}(0)=0.40$, $T_m=0.40$, $s=0.50$) and (b) $M=1$ with intersections in the second-order transition (SOT) regime ($h_{ex}(0)=0.35$, $T_m=1.40$, $s=0.50$), corresponding to Fig.~3(a) in the main text.  
  (c,d) Phase diagrams for (c) $M=0$ with alternative parameters ($h_{ex}(0)=0.65$, $T_m=0.45$, $s=0.50$) and (d) $M=1$ with intersections in the first-order transition (FOT) regime ($h_{ex}(0)=0.68$, $T_m=1.40$, $s=0.50$).  
  (e) Enlarged $h_{ex}(0)$-$T_m$ phase diagram for $s=0.50$, with the parameters for (a-d) marked by white symbols with black outlines. The dashed-dotted black line denotes the $T_m = T_{c0}$ boundary. Different behaviors are classified by the color scale (same convention used in subsequent figures).}
  \label{M_value_comparison}
\end{figure}

We systematically calculated the $h$-$\theta$ phase diagrams for different monotonic behaviors. For the $M=0$ and $M=1$ cases far from the $T_m = T_{c0}$ boundary, parameters distinct from those used in the main text were selected. The resulting phase diagrams are compared in Fig.~\ref{M_value_comparison}.

\begin{figure}[ht]
  \centering
  \includegraphics[width=0.9\linewidth]{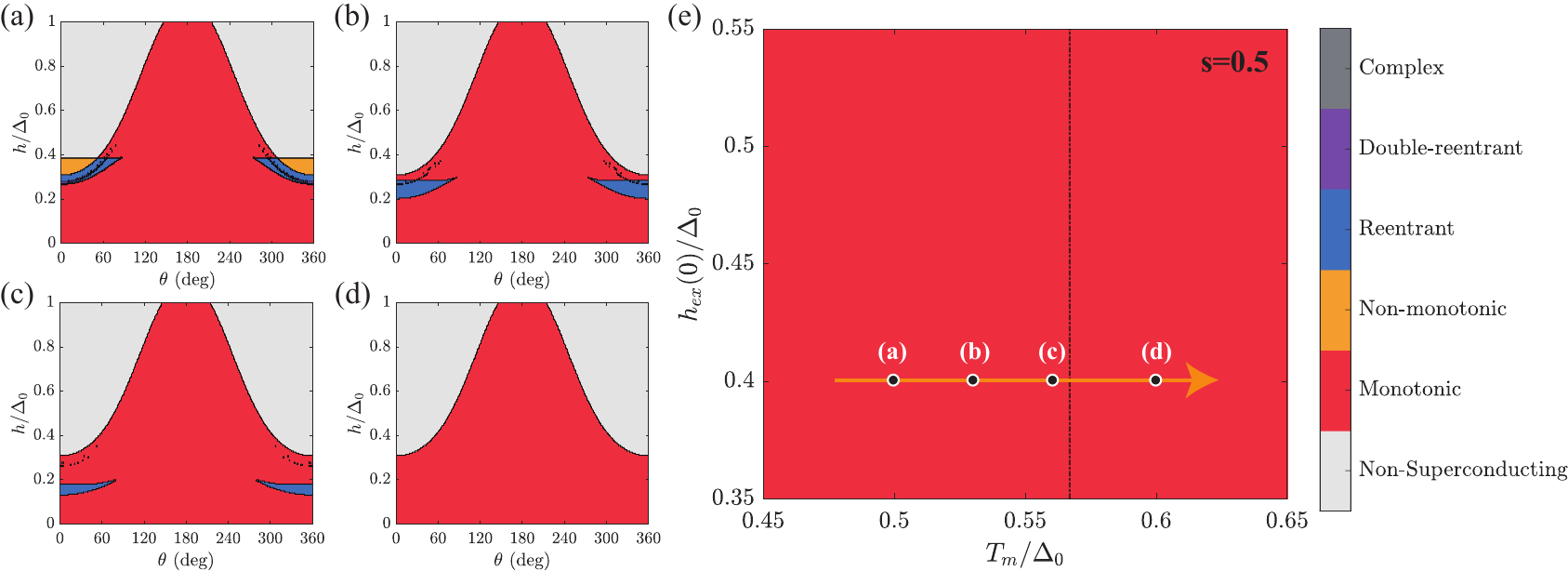}
  \caption{$h$-$\theta$ phase diagrams for four points across the $M=0$/$M=1$ boundary.  
  (a-d) Results for $M=0$ with $T_m=0.50$ (a), $0.53$ (b), $0.56$ (c), and $M=1$ with $T_m=0.60$ (d).  
  In all cases, $h_{ex}(0)=0.40$ and $s=0.50$.  
  (e) Enlarged $h_{ex}(0)$-$T_m$ phase diagram near the boundary, same as in Fig.~\ref{M_value_comparison}, with the parameters for (a-d) marked.}
  \label{M_across_boundary}
\end{figure}

According to the results in Fig.~\ref{M_value_comparison}, the magnetic response becomes topologically equivalent whenever the same intersection number $M$ is realized. It is therefore instructive to analyze the response near the $M=0$/$M=1$ boundary ($T_m = T_{c0}$ line). Figure~\ref{M_across_boundary} shows the $h$-$\theta$ phase diagrams for four representative points across this boundary.

By fixing $h_{ex}(0)=0.40$, the values of $T_m$ were varied from $0.50$ to $0.60$, crossing the boundary at $T_m \approx 0.57$. As seen in Fig.\ref{M_across_boundary}(a) and \ref{M_across_boundary}(d), the $T_m=0.50$ case is topologically equivalent to the $M=0$ regimes far from the boundary [Fig.\ref{M_value_comparison}(a,c)], while the $T_m=0.60$ case corresponds to $M=1$ regimes far from the boundary [Fig.\ref{M_value_comparison}(b,d)]. For the intermediate cases $T_m=0.53$ and $T_m=0.56$ [Figs.\ref{M_across_boundary}(b,c)], mixed magnetic responses emerge.

In summary, for $M=1$ cases sufficiently far from the $T_m=T_{c0}$ boundary, monotonic superconductivity persists under any external magnetic field until it is fully suppressed. In contrast, $M=0$ cases far from the boundary exhibit divergent responses, including monotonic $\rightarrow$ reentrant $\rightarrow$ non-monotonic sequences and related processes [Figs.~\ref{M_value_comparison}(a,c)]. Near the boundary, mixed transformations such as monotonic $\rightarrow$ reentrant $\rightarrow$ monotonic behavior can occur.

\newpage
\section{III. Magnetic-field-induced transformations for all phases}

In the main text, we presented magnetic-field-induced transformations only for monotonic behaviors. Here, we extend the analysis to other phases, with the results summarized in Table~2 of the main text.

\subsection{A. Non-monotonic phases}

As discussed in the main text [Fig.~1(c)], non-monotonic behavior can be further categorized into three distinct types. For the case of $M=2$ with both intersections lying in the SOT regime, the phase cannot occur when $s=0.50$. Therefore, to study all possible non-monotonic responses comprehensively, we set $s=1.0$ and calculated the corresponding $h$-$\theta$ phase diagrams.

\begin{figure}[ht]
  \centering
  \includegraphics[width=0.9\linewidth]{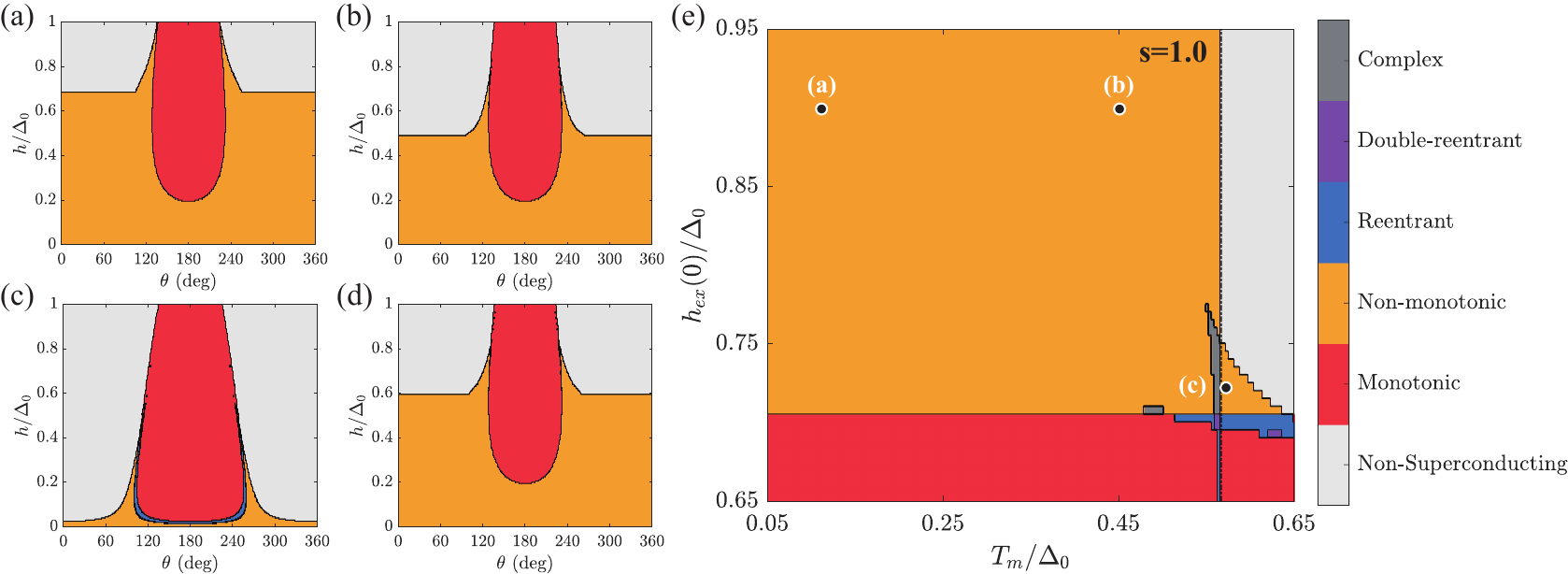}
  \caption{$h$-$\theta$ phase diagrams for different types of non-monotonic behavior.  
  (a) $M=1$ with intersections in the FOT regime ($h_{ex}(0)=0.90$, $T_m=0.10$, $s=1.0$).  
  (b) $M=1$ with intersections in the SOT regime ($h_{ex}(0)=0.90$, $T_m=0.45$, $s=1.0$).  
  (c) $M=2$ with both intersections in the SOT regime ($h_{ex}(0)=0.72$, $T_m=0.57$, $s=1.0$).  
  (d) $M=1$ with intersections in the FOT regime but close to the SOT boundary ($h_{ex}(0)=0.90$, $T_m=0.35$, $s=0.5$), shown for comparison with (b).  
  (e) Enlarged $h_{ex}(0)$-$T_m$ phase diagram for $s=1.0$, with parameters used in (a-c) marked.}
  \label{non-monotonic}
\end{figure}

As seen in Fig.~\ref{non-monotonic}, the magnetic response is primarily determined by the intersection number $M$. For $M=1$, the transformations are topologically equivalent, as illustrated in Figs.~\ref{non-monotonic}(a,b), and the distinction between FOT and SOT intersections is largely insignificant. In this case, the only possible transformation is non-monotonic $\rightarrow$ monotonic. For $M=2$, however, a narrow reentrant regime can appear, giving rise to more complex sequences such as non-monotonic $\rightarrow$ reentrant $\rightarrow$ monotonic, as shown in Fig.~\ref{non-monotonic}(c).

Additionally, to verify the universality of $M$-determined transformation behaviors, we also calculated the case of $s=0.50$ with $M=1$ in the FOT regime [Fig.\ref{non-monotonic}(d)]. Its response is topologically equivalent to Figs.\ref{non-monotonic}(a,b), further confirming this conclusion.

\subsection{B. Reentrant phases}

A similar analysis was carried out for reentrant behavior, which can also be categorized into three types according to their $h$-$\theta$ phase diagrams.

\begin{figure}[ht]
  \centering
  \includegraphics[width=0.9\linewidth]{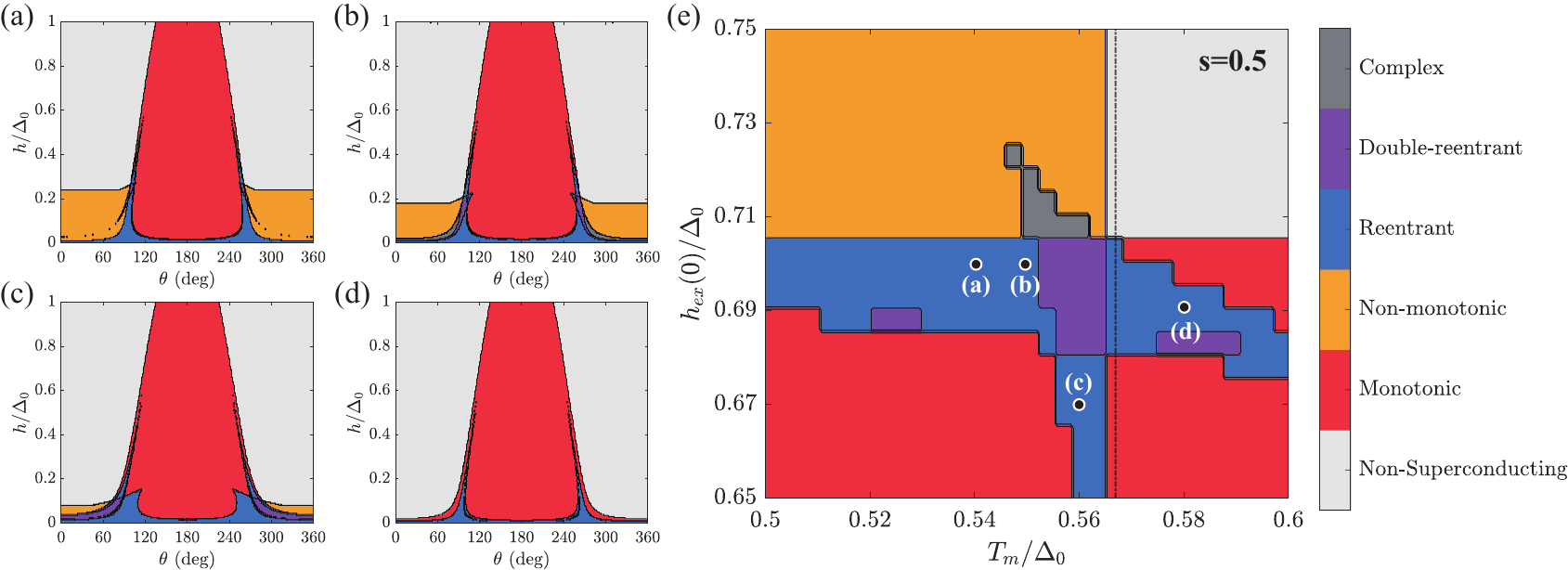}
  \caption{$h$-$\theta$ phase diagrams for different types of reentrant behavior.  
  (a) $M=2$ with both intersections in the FOT regime ($h_{ex}(0)=0.70$, $T_m=0.54$, $s=0.50$).  
  (b) $M=2$ with both intersections in the FOT regime but closer to the boundary ($h_{ex}(0)=0.70$, $T_m=0.55$, $s=0.50$).  
  (c) $M=2$ with both intersections in the SOT regime ($h_{ex}(0)=0.67$, $T_m=0.56$, $s=0.50$).  
  (d) $M=3$ with intersections spanning both FOT and SOT regimes ($h_{ex}(0)=0.69$, $T_m=0.58$, $s=0.50$).  
  (e) Enlarged $h_{ex}(0)$-$T_m$ phase diagram for $s=0.50$, with the parameters used in (a-d) marked.}
  \label{reentrant}
\end{figure}

Unlike the monotonic and non-monotonic cases, where magnetic responses are determined almost entirely by the intersection number $M$, the field-induced transformations of reentrant behaviors depend sensitively on the full topological relation between the phase transition lines and the $h_{ex}(T)$ curves, as illustrated in Figs.\ref{reentrant}(a-d). Cases with the same topology, such as Figs.\ref{reentrant}(a,b), are nearly equivalent, and the slight differences arise from the narrow parameter-space coverage of reentrant behavior, which makes them particularly sensitive to proximity to the topological boundary.

The greater diversity of responses in reentrant phases stems from their more complex topological relations, involving additional intersection points. The detailed transformation sequences are indicated in Fig.~\ref{reentrant} and are not elaborated here.

\subsection{C. Double-reentrant and complex phases}

For double-reentrant and complex behaviors, classification by topological relations is less practical because of their very narrow existence in the $h_{ex}(0)$-$T_m$ parameter space. Therefore, we present only representative $h$-$\theta$ phase diagrams, shown in Fig.~\ref{double_reentrant_complex}.

\begin{figure}[ht]
  \centering
  \includegraphics[width=1.0\linewidth]{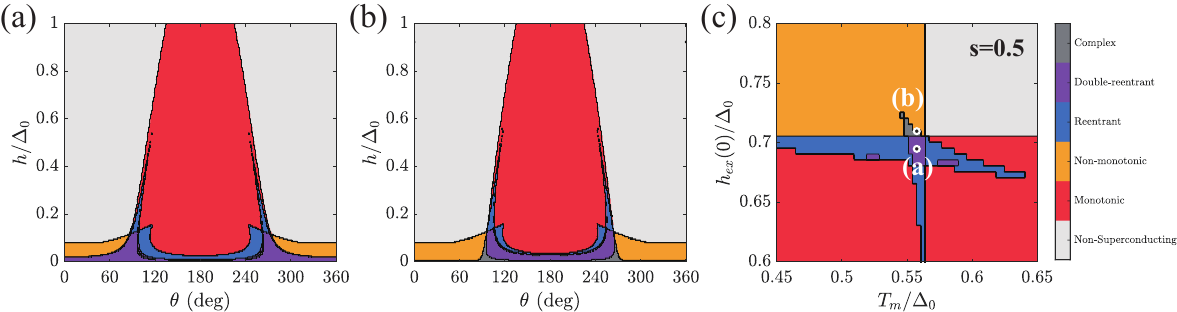}
  \caption{$h$-$\theta$ phase diagrams for representative double-reentrant and complex behaviors.  
  (a) Double-reentrant case with $h_{ex}(0)=0.69$, $T_m=0.56$, $s=0.50$.  
  (b) Complex case with $h_{ex}(0)=0.71$, $T_m=0.56$, $s=0.50$.
  (c) Enlarged $h_{ex}(0)$-$T_m$ phase diagram for $s=0.50$ near the double-reentrant and complex regimes, with parameters used in (a-b) marked.}
  \label{double_reentrant_complex}
\end{figure}

The magnetic responses of the double-reentrant and complex behaviors show strong similarities, closely resembling the field-induced transformations of the reentrant cases in Figs.\ref{reentrant}(a-d). This similarity arises from their proximity in parameter space. In principle, double-reentrant and complex phases could be further subdivided, but the results in Fig.\ref{double_reentrant_complex}, together with the above discussion, indicate that such classification would be of little significance.

Interestingly, by comparing Fig.\ref{reentrant} with Fig.\ref{double_reentrant_complex}(a), one finds that the double-reentrant phase lies at the center of three distinct reentrant regimes. Consequently, its magnetic response shares features with each reentrant type, even though the different reentrant phases themselves exhibit divergent transformations under external fields.

\subsection{D. Transformations of non-superconducting phases}

Most strikingly, non-superconducting states, where superconductivity is suppressed by a strong ferromagnetic exchange field, can be transformed into various superconducting phases through compensation of the ferromagnetism. Representative transformations under external magnetic fields are shown in Fig.~\ref{non_superconducting}.

\begin{figure}[ht]
  \centering
  \includegraphics[width=1.0\linewidth]{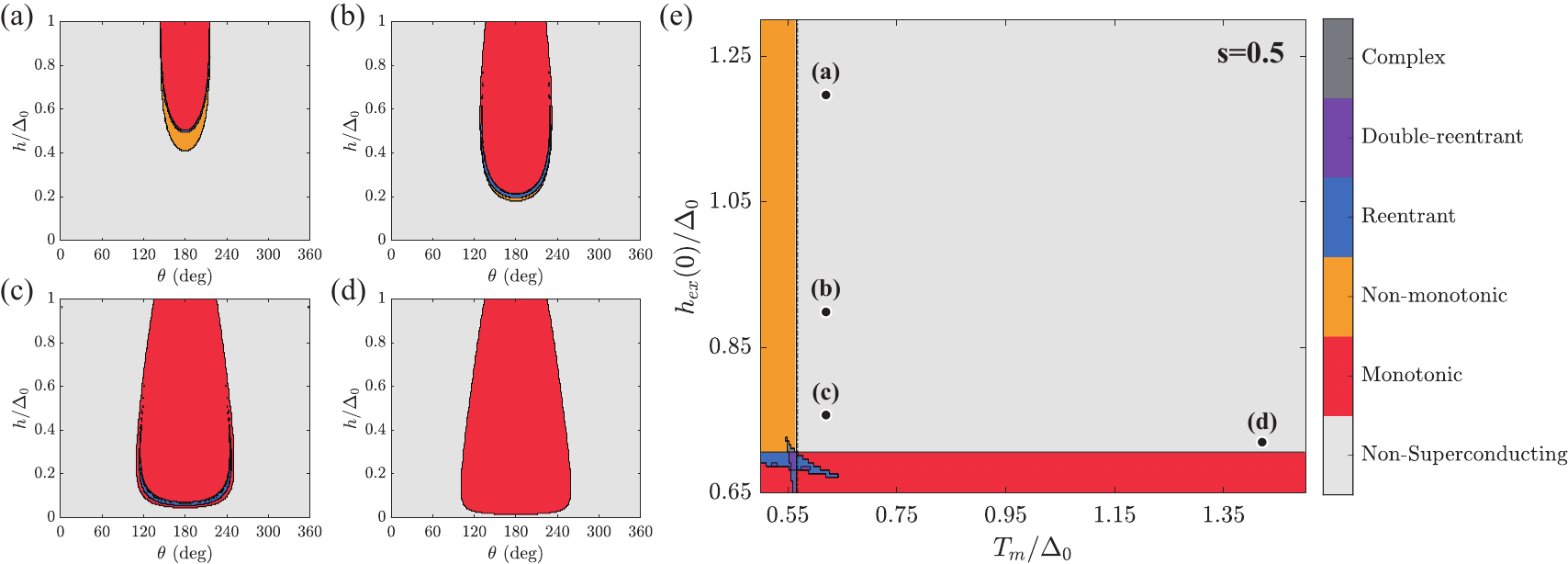}
  \caption{$h$-$\theta$ phase diagrams for representative non-superconducting states.  
  (a) $h_{ex}(0)=1.20$, $T_m=0.60$, $s=0.50$.  
  (b) $h_{ex}(0)=0.90$, $T_m=0.60$, $s=0.50$.  
  (c) $h_{ex}(0)=0.75$, $T_m=0.60$, $s=0.50$.  
  (d) $h_{ex}(0)=0.72$, $T_m=1.40$, $s=0.50$.  
  (e) Enlarged $h_{ex}(0)$-$T_m$ phase diagram for $s=0.50$ including the non-superconducting regimes, with the parameters of (a-d) marked.}
  \label{non_superconducting}
\end{figure}

As shown in Fig.\ref{non_superconducting}(a-d), the magnetic responses of non-superconducting states can be classified into three types. The first type lies in the upper-left region of the $h_{ex}(0)$-$T_m$ phase diagram, where $h_{ex}(0) \gg H_p$ but $T_m \approx T_{c0}$ [Fig.\ref{non_superconducting}(a)]. The second type appears in the lower-right region, where $h_{ex}(0) \approx H_p$ but $T_m \gg T_{c0}$ [Fig.\ref{non_superconducting}(d)]. The third type occurs near the double-reentrant and complex regimes, where $h_{ex}(0) \approx H_p$ and $T_m \approx T_{c0}$, effectively acting as an intermediate between the first two types [Figs.\ref{non_superconducting}(b,c)]. For cases with both large $h_{ex}(0)$ and $T_m$, the response is less conclusive and may belong to one of these three categories.

Because magnetic-field-induced superconductivity emerging from a non-superconducting state corresponds to the so-called high-field reentrant superconductivity (HFRS), it is of particular interest to analyze its magnetic response in more detail, given the relevance to recent HFRS experiments. By fixing $\theta=180^\circ$, the calculated $\Delta(T,h)$ relations for the non-superconducting cases discussed above are shown in Fig.~\ref{non_superconducting_h_t}.

\begin{figure}[ht]
  \centering
  \includegraphics[width=1.0\linewidth]{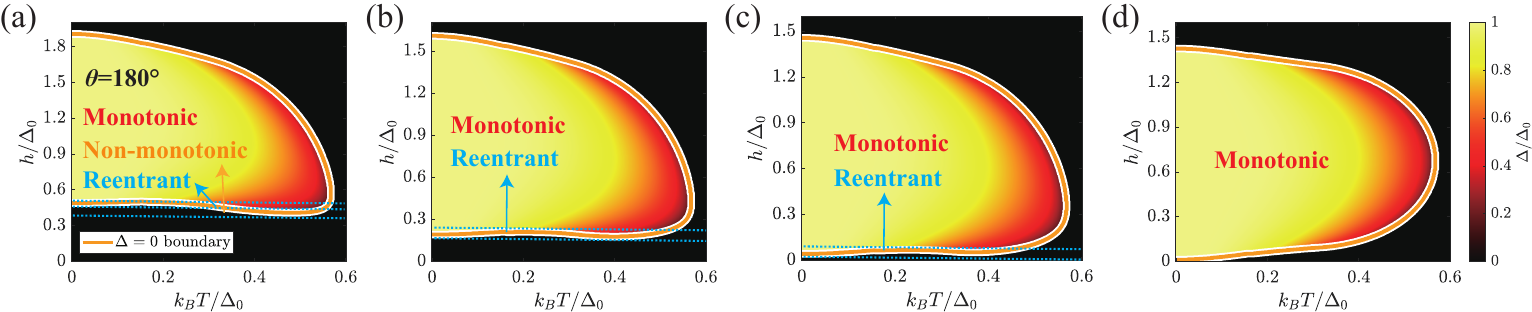}
  \caption{Magnetic-field-induced transformations of non-superconducting states at $\theta=180^\circ$.  
  (a-d) $\Delta(T,h)$ maps corresponding to the cases in Fig.~\ref{non_superconducting}(a-d). Each distinct behavior is labeled (very narrow regimes are omitted). The yellow contour denotes the $\Delta=0$ boundary, i.e., the $h$-$T$ phase transition line (same convention used in subsequent figures). The gap amplitude is indicated by the color scale.}
  \label{non_superconducting_h_t}
\end{figure}

In Figs.~\ref{non_superconducting_h_t}(a-d), the yellow contour separates superconducting and non-superconducting regions and can be regarded as the $h$-$T$ phase transition line. Notably, in Fig.~\ref{non_superconducting_h_t}(a), the transition curve is asymmetric with respect to $h_{ex}(0)$, consistent with recent experiments on orphan $\mathrm{UTe_2}$ \cite{frank2024orphan}, but incompatible with the conventional WHH theory \cite{rubi2025extreme}, which requires symmetry about $h_{ex}(0)$.  

Additionally, as the parameters evolve from Fig.\ref{non_superconducting}(a) to Fig.\ref{non_superconducting}(d), the $h$-$T$ phase transition lines also shift correspondingly in Fig.~\ref{non_superconducting_h_t}(a-d). These lines gradually become symmetric about $h_{ex}(0)$, which can be understood as a consequence of $h_{ex}(T)$ remaining nearly constant within $T \in [0,T_{c0}]$. Thus, by appropriately tuning the parameters, the slight asymmetry in the $h$-$T$ phase diagram can be controlled, consistent with recent experimental observations \cite{rubi2025extreme}.

A more detailed analysis of the $h$-$T$ phase transition line in high-field reentrant superconductivity (HFRS) will be presented in a future work and lies beyond the scope of this article and Supplementary Material.

In conclusion, we have shown that the magnetic response of monotonic and non-monotonic behaviors is primarily governed by the intersection number $M$, whereas field-induced transformations of reentrant behaviors are controlled by the full topological relation between the phase transition lines and the $h_{ex}(T)$ curve. The detailed transformation processes are presented in Figs.~\ref{M_value_comparison}-\ref{double_reentrant_complex}. In principle, $h$-$T$ phase diagrams can be computed for any $\theta$, but these are omitted here due to space limitations. We also note that the physics near topological boundaries is especially intriguing and will be addressed in future work. Finally, we have discussed the transformation from non-superconducting to superconducting phases, which shows good agreement with recent experimental results on HFRS.

\newpage
\section{IV. Comparisons with experimental observations}

In the main text, we demonstrated that our calculations can account for nearly all reported experimental observations. The three temperature-dependent superconducting behaviors are well established and have been widely observed in $R$-$T$ measurements \cite{eisaki1994competition,huy2007superconductivity,ishikawa1977destruction,fertig1977destruction,tran2024reentrant,jiang2009superconductivity,paramanik2013reentrant,anisotropic1996}. These behaviors are naturally reproduced within our framework and further classified, as shown in Fig.~1 of the main text. Because these results are already extensively documented, we did not reproduce the specific $R$-$T$ curves here, and readers are referred to the cited works for detailed data\footnote{Representative $R$-$T$ curves for each type include: monotonic ($\mathrm{ErNi_2B_2C}$, Fig.~2 of \cite{eisaki1994competition}); non-monotonic ($\mathrm{ErRh_4B_4}$, Fig.~1 of \cite{fertig1977destruction}); reentrant ($\mathrm{HoNi_2B_2C}$, Fig.~2 of \cite{eisaki1994competition}).}. 

To provide more direct support for our theory, we next discuss several detailed experimental results in greater depth.

\subsection{A. Specific heat and tunneling measurements in $\mathbf{EuRh_4B_4}$}

Our calculations indicate that the superconducting transition should be first order at low temperatures but second order at higher temperatures. This distinction can be directly probed through specific-heat experiments. In addition, the calculated $\Delta(T)$ relation provides more detailed information that can be compared with tunneling measurements, which remain relatively scarce. The relevant comparisons are summarized below.  

Particular attention is given to $\mathrm{EuRh_4B_4}$, the only system for which tunneling data are available \cite{crespo2009evolution}. In transport experiments\footnote{See Fig.~4 of \cite{fertig1977destruction} and Fig.~1 of \cite{woolf1979superconducting}.}, a divergent low-temperature specific-heat anomaly was observed \cite{fertig1977destruction}, consistent with a first-order transition, while a discontinuity in the high-temperature regime confirmed the presence of a second-order transition. Notably, $\mathrm{LuRh_4B_4}$ shows only a high-temperature discontinuity, again supporting this interpretation [Fig.~1(a) in the main text]. Most strikingly, the temperature dependence of the superconducting gap extracted from tunneling spectroscopy\footnote{See Fig.~2(c) of \cite{crespo2009evolution}.} exhibits remarkable agreement with our calculated results, as shown in Fig.~\ref{Specific_heat_tunnelling}.

\begin{figure}[ht]
  \centering
  \includegraphics[width=0.6\linewidth]{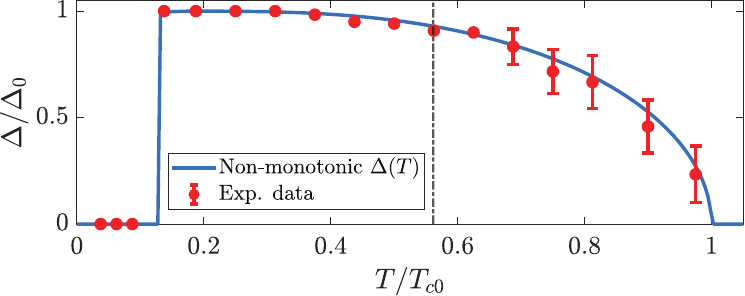}
  \caption{Comparison between the calculated $\Delta(T)$ relation and tunneling measurements of $\mathrm{EuRh_4B_4}$. Experimental data were extracted from \cite{crespo2009evolution}. The dashed-dotted black line marks the critical point separating the FOT and SOT regimes. Parameter set: $h_{ex}(0)=0.90$, $T_m=0.10$, $s=0.50$.}
  \label{Specific_heat_tunnelling}
\end{figure}

Despite the excellent consistency, only limited tunneling measurements are currently available for comparison. We are therefore particularly enthusiastic about future experiments that could provide more detailed data for further validation.

\subsection{B. Magnetic response in $\mathbf{HoNi_2B_2C}$, $\mathbf{TmNi_2B_2C}$, and other FSCs}

The magnetic responses of the various phases were systematically analyzed in Section~III. The $h$-$\theta$ phase diagrams from Figs.~\ref{M_value_comparison}-\ref{non_superconducting} can account for all experimentally observed field-induced transformations. Here we highlight several representative monotonic cases.  

For $\mathrm{ErNi_2B_2C}$, the superconductivity remains monotonic under magnetic field until it is completely suppressed, consistent with the $h$-$\theta$ phase diagram of Fig.~\ref{M_value_comparison}(d). $\mathrm{TmNi_2B_2C}$ exhibits a transformation from monotonic to non-monotonic behavior, which is well explained by Fig.~\ref{M_value_comparison}(c); the absence of reentrant behavior can be attributed to its narrow stability region. In contrast, $\mathrm{HoNi_2B_2C}$ shows monotonic $\rightarrow$ reentrant transformations under field, consistent with the $h$-$\theta$ phase diagrams of Figs.~\ref{M_across_boundary}(b,c), where $T_m \approx T_{c0}$.  
To further compare with experimentally obtained $H$-$T$ phase diagrams\footnote{See experimental results for $\mathrm{ErNi_2B_2C}$ and $\mathrm{TmNi_2B_2C}$ in Fig.~3 of \cite{eisaki1994competition}, and for $\mathrm{HoNi_2B_2C}$ in Fig.~8 of \cite{anisotropic1996}.}, we have carried out quantitative calculations. A more detailed comparison, including orbital depairing and additional effects, will be presented in future work.

\begin{figure}[ht]
  \centering
  \includegraphics[width=0.9\linewidth]{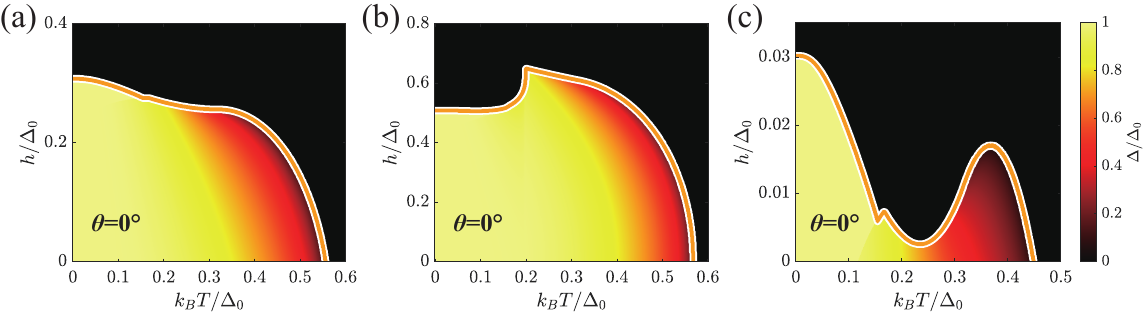}
  \caption{Representative $H$-$T$ phase diagrams ($\theta=0^\circ$) for FSCs with monotonic temperature dependence.  
  (a) $\mathrm{ErNi_2B_2C}$-type: $h_{ex}(0)=0.20$, $T_m=0.20$, $s=0.50$.  
  (b) $\mathrm{TmNi_2B_2C}$-type: $h_{ex}(0)=0.40$, $T_m=0.58$, $s=0.50$.  
  (c) $\mathrm{HoNi_2B_2C}$-type: $h_{ex}(0)=0.677$, $T_m=0.58$, $s=0.50$.}
  \label{h_t_comparison}
\end{figure}

The calculated $H$-$T$ phase-transition lines show good agreement with experimental results in both shape and trend\footnote{See magnetic-field-dependent $R$-$T$ curves for $\mathrm{ErNi_2B_2C}$ and $\mathrm{TmNi_2B_2C}$ in Fig.~2 of \cite{eisaki1994competition}, and for $\mathrm{HoNi_2B_2C}$ in Figs.~3 and 4 of \cite{anisotropic1996}.}.  

\subsection{C. Explanation of divergent behaviors in the same material}

Experiments have shown that the same material may exhibit slightly different temperature-dependent behaviors. For example, $\mathrm{HoNi_2B_2C}$ has been reported to display both monotonic and reentrant behavior\footnote{See reentrant $R$-$T$ curves in Fig.~1 of \cite{eisaki1994competition}, and monotonic $R$-$T$ curves in Fig.~2 of \cite{anisotropic1996}.}. These divergent results can be naturally explained within the $h_{ex}(0)$-$T_m$ phase diagram. It has been established that $T_m \approx T_{c0}$ in $\mathrm{HoNi_2B_2C}$ \cite{hu2025lithium}, placing the system near the phase boundary separating monotonic and reentrant regimes, as illustrated in Fig.~\ref{divergent_explanation}(d).

\begin{figure}[ht]
  \centering
  \includegraphics[width=1.0\linewidth]{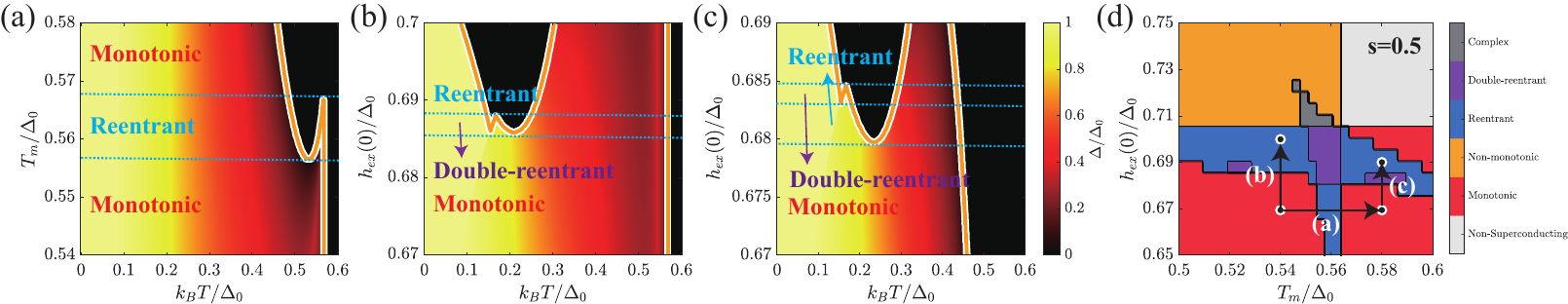}
  \caption{Calculated $\Delta(T)$ relations for varying $h_{ex}(0)$ and $T_m$ with $s=0.50$.  
  (a) $T_m=0.54 \rightarrow 0.58$ with fixed $h_{ex}(0)=0.67$. 
  (b) $h_{ex}(0)=0.67 \rightarrow 0.70$ with fixed $T_m=0.54$. 
  (c) $h_{ex}(0)=0.67 \rightarrow 0.69$ with fixed $T_m=0.58$. 
  (d) Enlarged $h_{ex}(0)$-$T_m$ phase diagram for $s=0.50$ near the monotonic-reentrant boundary. The parameter variations used in (a-c) are indicated by arrows.}
  \label{divergent_explanation}
\end{figure}

To further demonstrate how slight parameter variations can drastically alter the temperature-dependent superconducting behavior, we analyzed several cases near this boundary [Figs.~\ref{divergent_explanation}(a-c)]. A monotonic state can be tuned into a reentrant state, and even into a double-reentrant state (though the latter is unlikely due to its extremely narrow stability region), with parameter changes as small as $1\%$. Such sensitivity likely explains the divergent behaviors reported in experiments.

\newpage
\section{V. Fragile $\mathbf{T_c}$ enhancement and orbital depairing effects}

In the main text, we showed that $T_c$ can be enhanced either slightly or significantly, depending on the circumstances. Experimentally, however, only slight enhancements have been reported. For example, in lithium-intercalated FeSe, the observed $T_c$ increase was only about $0.5\%$ and proved highly fragile with respect to field tilting \cite{hu2025lithium}. This angular sensitivity can be understood in terms of orbital depairing. Because the sample thickness ($14\,\mathrm{nm}$) is much smaller than the coherence length, the in-plane field can be treated as providing only an effective exchange field, with orbital (vortex) depairing largely excluded \cite{hu2025lithium}. By contrast, an out-of-plane field inevitably introduces orbital depairing, leading to additional suppression of superconductivity. As a result, the observed $T_c$ enhancement is fragile under tilted magnetic fields.

Here, we first show that the magnitude of the slight $T_c$ enhancement can be tuned by selecting appropriate values of $h_{ex}(0)$ and $T_m$ (with $s=0.50$). As expected, the enhancement decreases with smaller $h_{ex}(0)$, since it originates from the compensation of the exchange field. Assuming orbital effects are negligible (as in the case of in-plane magnetic fields), we calculate the angular dependence of $T_c$ enhancement for different parameter sets.

\begin{figure}[ht]
  \centering
  \includegraphics[width=0.73\linewidth]{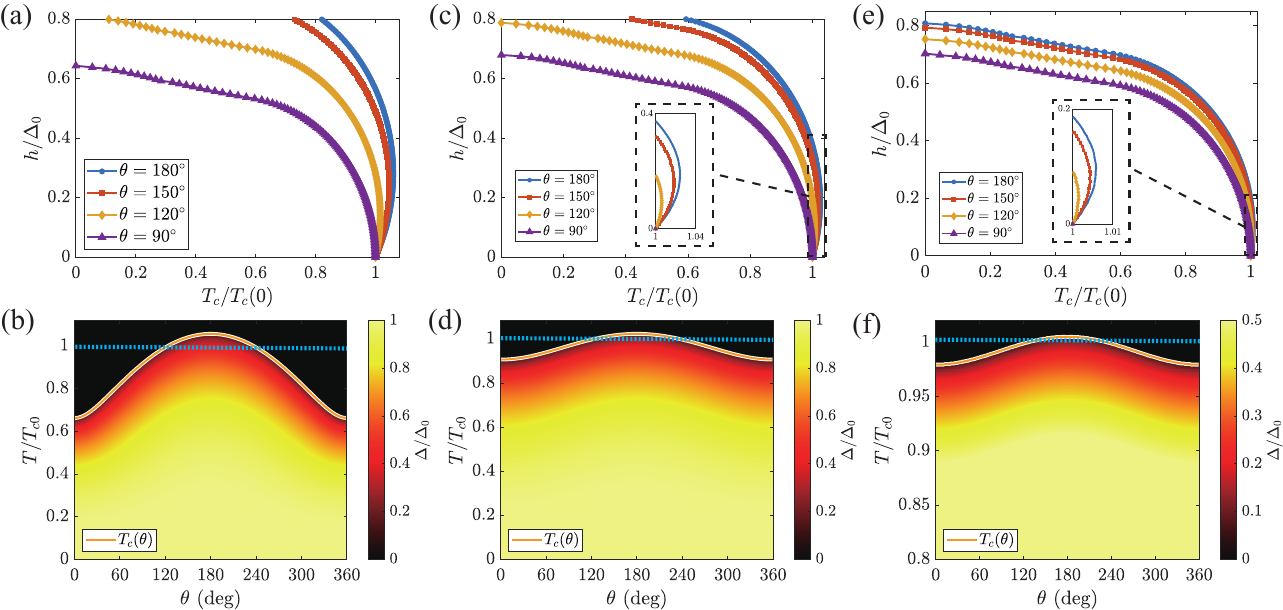}
  \caption{Magnetic-field-induced $T_c$ enhancement for different parameters without orbital depairing.  
  (a,c,e) Field dependence of $T_c$ along $\theta=90^{\circ}, 120^{\circ}, 150^{\circ}, 180^{\circ}$ for three cases: relatively large (a-b), moderate (c-d), and small (e-f) enhancements. Insets: zoomed views highlighting the slight enhancements.  
  (b,d,f) $\Delta(T,\theta)$ maps at fixed $h$ values chosen to maximize $T_c$ ($h=0.30$, $0.20$, $0.10$ for b, d, f respectively). Blonde curves trace the field-enhanced $T_c$ as a function of $\theta$, while light-blue lines denote the zero-field $T_c(0)$.  
  Parameters: (a-b) $h_{ex}(0)=0.30$, $T_m=1.40$, $s=0.50$; (c-d) $h_{ex}(0)=0.20$, $T_m=1.40$, $s=0.50$; (e-f) $h_{ex}(0)=0.10$, $T_m=1.40$, $s=0.50$.}
  \label{slight_enhancement}
\end{figure}

As shown in Fig.\ref{slight_enhancement}(a), (c), and (e), $T_c$ enhancement can be achieved for $90^{\circ}<\theta<270^{\circ}$ when orbital effects are excluded. This behavior is readily understood as compensation of the total exchange field. Furthermore, when $h$ is fixed to maximize $T_c$ (as in the main text), the enhancement is restricted to $120^{\circ}<\theta<240^{\circ}$ [Fig.\ref{slight_enhancement}(b), (d), and (f)], again consistent with the compensation picture. These results are in full agreement with the main-body analysis. Importantly, as illustrated in Figs.~\ref{slight_enhancement}(e,f), the enhancement can become extremely slight, rendering it highly vulnerable to perturbations.

To further demonstrate the fragility of $T_c$ enhancement, we next include orbital effects. Within the Ginzburg-Landau (GL) framework, free-energy density $f_s$ of a superconductor can be expressed as

\begin{equation}\label{GL_thoery}
    f_s = f_0 + \alpha |\psi|^2 + \frac{\beta}{2}|\psi|^4 
    + \frac{1}{2m^{*}}\left| \left(-i\hbar\nabla - 2e \mathbf{A}\right)\psi \right|^2,
\end{equation}

\noindent where $\psi$ is the superconducting order parameter ($\psi\sim \Delta$), and the orbital depairing effect is described by the last term in Eq.~(\ref{GL_thoery}). The orbital depairing energy under a magnetic field can be obtained from the first GL equation, which in this case is mathematically equivalent to solving the Schrödinger equation with a harmonic potential, a standard textbook result \cite{tinkham2004introduction}. Consequently, the orbital energy spectrum is quantized into Landau levels

\begin{equation}
    E_n = 2\!\left(n+\tfrac{1}{2}\right)\mu_B H_{\perp},
\end{equation}

\noindent where $\mu_B$ is the Bohr magneton and $H_{\perp}$ is the out-of-plane component of the external magnetic field. For the ground state, the orbital energy reduces to $E=\mu_B H_{\perp}$. Considering the averaged thermodynamic potential of the system and applying Eq.~(\ref{GL_thoery}), the reduced orbital contribution can be expressed as

\begin{equation}\label{orbit_free_energy}
    f_{orb} = \alpha \frac{\mu_BH_{\perp}}{\Delta_0} \left(\frac{\Delta}{\Delta_0}\right)^2,
\end{equation}

\noindent where $\alpha$ is a phenomenological coefficient chosen as $\alpha=1$ in our calculations. We note that this treatment is approximate, but it captures the essential depairing effect and illustrates the fragility of the $T_c$ enhancement.  

Eq.~(\ref{orbit_free_energy}) is incorporated into Eq.~(\ref{eq:fx}) to account for orbital depairing, and the same minimization procedure is then applied to determine $\Delta$. The angular dependence of $T_c$ enhancement with orbital depairing included was subsequently calculated for different parameter sets.

\begin{figure}[ht]
  \centering
  \includegraphics[width=0.73\linewidth]{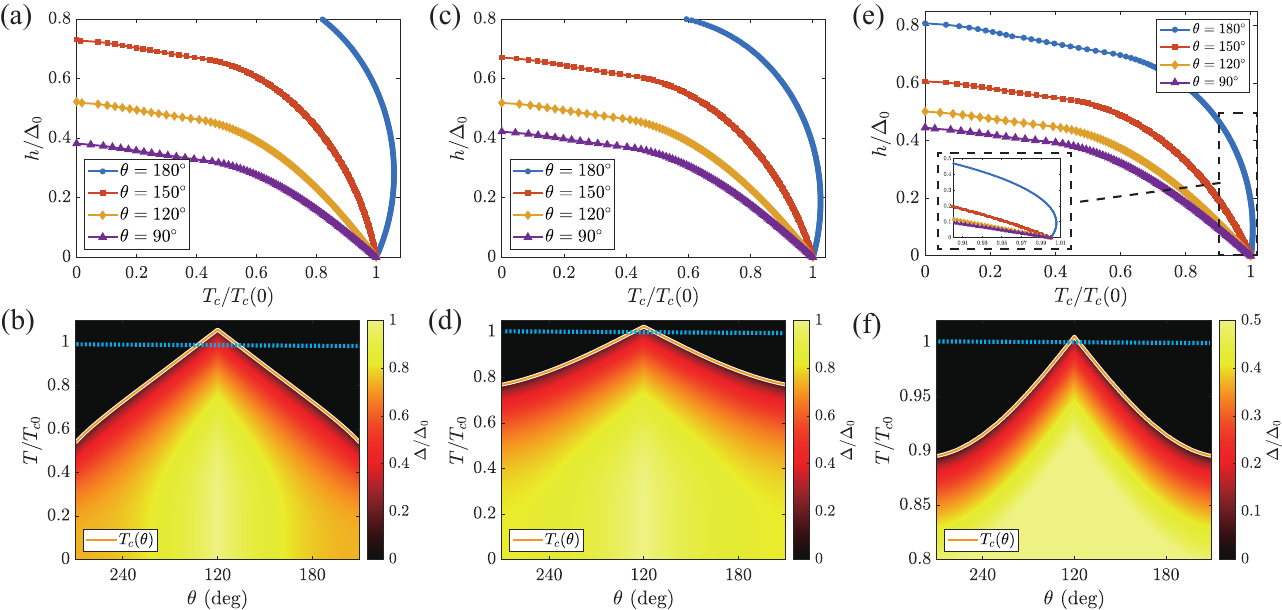}
  \caption{Magnetic-field-induced $T_c$ enhancement with orbital depairing included.  
  (a,c,e) Field dependence of $T_c$ along $\theta=90^{\circ}, 120^{\circ}, 150^{\circ}, 180^{\circ}$ for three cases: relatively large (a-b), moderate (c-d), and small (e-f) enhancements.  
  (b,d,f) $\Delta(T,\theta)$ maps at fixed $h$ chosen to maximize $T_c$. Blonde curves trace the field-enhanced $T_c$ as a function of $\theta$, while light-blue lines denote the zero-field $T_c(0)$.  
  Parameters are identical to those used in Fig.~\ref{slight_enhancement} for direct comparison.}
  \label{slight_enhancement_with_orbits}
\end{figure}

By comparing Figs.~\ref{slight_enhancement} and \ref{slight_enhancement_with_orbits}, it is clear that orbital effects are detrimental to $T_c$ enhancement. For $\theta=180^\circ$ (in-plane magnetic field), the curves remain unchanged, consistent with the assumption that orbital effects are negligible in sufficiently thin films for in-plane fields. At other angles, however, $T_c$ decreases rapidly as $\theta$ shifts away from $180^\circ$ toward $90^\circ$, highlighting the fragility of the enhancement once orbital depairing is included. Moreover, the results in Figs.~\ref{slight_enhancement_with_orbits}(c) and \ref{slight_enhancement_with_orbits}(d) show qualitative agreement with experimental data\footnote{See Fig.~3(d), 3(e) and 3(f) of \cite{hu2025lithium}.}. 

The angular sensitivity is further highlighted by calculations with finer angular resolution of $2^\circ$ near $\theta=180^\circ$, as shown in Fig.~\ref{angular_sensitivity}. It is found that the $T_c$ enhancement vanishes within $5^\circ$ of misalignment, confirming its fragility under orbital-depairing perturbations.

\begin{figure}[ht]
  \centering
  \includegraphics[width=0.3\linewidth]{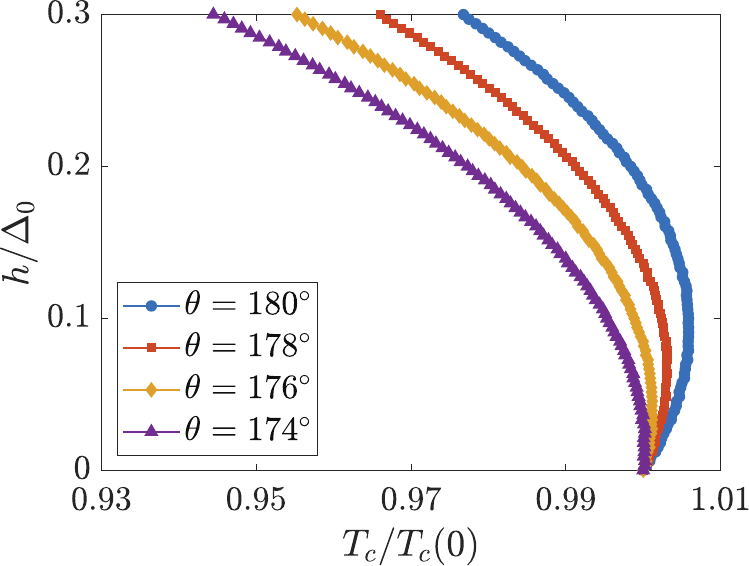}
  \caption{Calculated $T_c$ with fine angular resolution ($\theta=174^\circ$-$180^\circ$), illustrating the extreme sensitivity of the enhancement to field orientation.}
  \label{angular_sensitivity}
\end{figure}

\end{widetext}


\begin{thebibliography}{40}
\bibitem [{myN()}]{myNote}%
  \BibitemOpen
  \href@noop {} {}\bibinfo {note} {The orbital effect is neglected in the
  present discussion. However, it may be practically important in materials
  with a small Maki parameter. Corresponding calculations will be presented in
  future work, where inclusion of orbital effects is expected to yield improved
  quantitative agreement with experimental results.}\BibitemShut {Stop}%
\end{thebibliography}
\end{document}